\let\ifarxiv=\iftrue     
\documentclass[12pt,a4paper]{article}

\ifarxiv\ifnum\pdfoutput=1\else
\PassOptionsToPackage{hypertex}{hyperref}
\PassOptionsToPackage{draft}{graphicx}
\fi\fi

\usepackage{amsmath,amssymb}
\usepackage[bookmarks=true,hyperfigures=true]{hyperref}
\usepackage{graphicx}
\usepackage[nosort]{cite}
\usepackage[bulletsep]{collref}
\usepackage{feynmp}
\usepackage{multirow}
\usepackage{listings}
\usepackage{longtable}
\usepackage{multicol}
\usepackage{hhline}

\usepackage[a4paper,text={173mm,216mm},centering]{geometry}

\allowdisplaybreaks[3]

\numberwithin{equation}{section}

\usepackage[font=small,labelfont=bf,width=0.85\textwidth]{caption}

\makeatletter
\let\old@startsection=\@startsection
\renewcommand{\@startsection}[6]{\old@startsection{#1}{#2}{#3}{#4}{#5}{#6\mathversion{bold}}}
\makeatother

\makeatletter
\newlength{\apb@width}
\newcommand{\autoparbox}[2][c]{\settowidth{\apb@width}{#2}\parbox[#1]{\apb@width}{#2}}

\makeatother

\let\oldPhi=\Phi
\let\oldPsi=\Psi
\let\oldGamma=\Gamma
\let\oldDelta=\Delta
\let\oldSigma=\Sigma
\let\oldLambda=\Lambda
\let\oldTheta=\Theta
\let\oldPi=\Pi
\let\oldXi=\Xi
\let\oldUpsilon=\Upsilon
\let\oldOmega=\Omega
\renewcommand{\Phi}{\mathnormal{\oldPhi}}
\renewcommand{\Psi}{\mathnormal{\oldPsi}}
\renewcommand{\Gamma}{\mathnormal{\oldGamma}}
\renewcommand{\Sigma}{\mathnormal{\oldSigma}}
\renewcommand{\Delta}{\mathnormal{\oldDelta}}
\renewcommand{\Theta}{\mathnormal{\oldTheta}}
\renewcommand{\Lambda}{\mathnormal{\oldLambda}}
\renewcommand{\Pi}{\mathnormal{\oldPi}}
\renewcommand{\Xi}{\mathnormal{\oldXi}}
\renewcommand{\Upsilon}{\mathnormal{\oldUpsilon}}
\renewcommand{\Omega}{\mathnormal{\oldOmega}}

\ifx\genfrac\sdflkaj\else\fi

\newcommand{\vev}[1]{\langle#1\rangle}
\newcommand{\bev}[1]{ [#1]}
\newcommand{\pslash}{\slash\!\!\! p}

\newcommand{\bra}[1]{\langle #1 \rvert}

\newcommand{\cket}[1]{|#1]}

\newcommand{\be}{\begin{equation}}
\newcommand{\ee}{\end{equation}}

\newcommand{\Tr}{\mathop{\mathrm{Tr}}}
\newcommand{\sign}{\mathop{\mathrm{sign}}}

\newcommand{\nn}{\nonumber}

\makeatletter
\def\mr@ignsp#1 {\ifx\:#1\@empty\else #1\expandafter\mr@ignsp\fi}%
\newcommand{\multiref}[1]{\begingroup
\xdef\mr@no@sparg{\expandafter\mr@ignsp#1 \: }%
\def\mr@comma{}%
\@for\mr@refs:=\mr@no@sparg\do{\mr@comma\def\mr@comma{,}\ref{\mr@refs}}%
\endgroup}
\makeatother

\ifx\href\asklfhas\newcommand{\href}[2]{#2}\fi

\newcommand{\hypref}[2]{\ifx\href\asklfhas #2\else\href{#1}{#2}\fi}

\newcommand{\secref}[1]{Sec.~\multiref{#1}}

\renewcommand{\eqref}[1]{(\multiref{#1})}
\newcommand{\eqn}[1]{eq.~\eqref{#1}}
\def\Eqn#1{Equation~(\ref{#1})}

\renewcommand{\i}{\ensuremath{\mathrm{i}}}

\def\ib{{\bar\imath}}
\def\tree{{\rm tree}}

\newcommand{\indexfett}[2][f]{\if#1f{\index{#2|textbf}}\else{\index{#2}}\fi}

\ifx\hypersetup\sadfkjashdfkxja\newcommand\hypersetup[1]{}\newcommand{\texorpdfstring}[2]{#1}\fi

\hypersetup{plainpages=false}
\hypersetup{pdfpagemode=UseNone}
\hypersetup{bookmarksnumbered=true}
\hypersetup{pdfstartview=FitH}
\hypersetup{colorlinks=false}
\hypersetup{citebordercolor={.5 1 .5}}
\hypersetup{urlbordercolor={.5 1 1}}
\hypersetup{linkbordercolor={1 .7 .7}}

\renewcommand{\=}{\mathrel{\phantom{=}}}

\newcommand{\ang}[2]{\langle #1\;#2\rangle}
\newcommand{\bracket}[2]{\langle #1 \vert #2 \rangle}

\newcommand{\+}{\negthinspace+\negthinspace}

\allowdisplaybreaks

\begin{document}
\thispagestyle{empty}

\begin{flushright}
CERN-PH-TH/2010-230\\
SLAC--PUB--14278\\
HU-EP-10/57
\end{flushright}

\begingroup\centering
{\Large\bfseries\mathversion{bold}
All tree-level amplitudes in massless QCD\par}%
\hypersetup{pdfsubject={}}%
\hypersetup{pdfkeywords={}}%
\ifarxiv\vspace{15mm}\else\vspace{15mm}\fi

\begingroup\scshape\large 
Lance J. Dixon,
\endgroup
\vspace{5mm}

\begingroup\ifarxiv\small\fi
\textit{
Theory Group, Physics Department, \\
CERN, CH--1211 Geneva 23, Switzerland\\ 
\vspace{0.35cm}
and\\ 
\vspace{0.35cm}
SLAC National Accelerator Laboratory, \\
Stanford University, Stanford, CA 94309, USA
 }\par
\texttt{lance@slac.stanford.edu\phantom{\ldots}}
\endgroup

\vspace{1cm}

\begingroup\scshape\large 
Johannes M. Henn, Jan Plefka and Theodor Schuster
\endgroup
\vspace{5mm}

\begingroup\ifarxiv\small\fi
\textit{Institut f\"ur Physik, Humboldt-Universit\"at zu Berlin, \\
Newtonstra{\ss}e 15, D-12489 Berlin, Germany}\\[0.2cm]
\ifarxiv\texttt{\{henn,plefka,theodor\}@physik.hu-berlin.de\phantom{\ldots}}\fi
\endgroup
\vspace{1cm}

\textbf{Abstract}\vspace{5mm}\par
\begin{minipage}{14.7cm}
We derive compact analytical formulae for all tree-level
color-ordered gauge theory amplitudes involving any number of external
gluons and up to four massless quark-anti-quark pairs. A general formula is
presented based on the combinatorics of paths along a rooted tree
and associated determinants. Explicit expressions are displayed
for the next-to-maximally helicity violating (NMHV) and
next-to-next-to-maximally helicity violating (NNMHV) gauge theory
amplitudes.
Our results are obtained by projecting the previously-found expressions
for the super-amplitudes of the maximally supersymmetric
super Yang-Mills theory (${\cal N}=4$ SYM) onto the relevant components
yielding all gluon-gluino
tree amplitudes in ${\cal N}=4$ SYM. We show how
these results carry over to the corresponding QCD amplitudes, including
massless quarks of different flavors as well as a single electroweak
vector boson.  The public {\tt Mathematica} package {\tt GGT} 
is described, which encodes the results of this work and yields 
analytical formulae for all ${\cal N}=4$ SYM gluon-gluino trees.
These in turn yield all QCD trees with up to four external
arbitrary-flavored massless quark-anti-quark pairs.
\end{minipage}\par
\endgroup
\newpage


\setcounter{tocdepth}{2}
\hrule height 0.75pt
\tableofcontents
\vspace{0.8cm}
\hrule height 0.75pt
\vspace{1cm}

\setcounter{tocdepth}{2}


\section{Introduction and conclusions}

Scattering amplitudes play a central role in gauge theory.
At a phenomenological level, they are critical to the prediction
of cross sections at high-energy colliders, for processes within and
beyond the Standard Model.  Efficient evaluation of scattering amplitudes
involving many quarks and gluons is particularly important at machines such
as the Large Hadron Collider (LHC), in which complex, multi-jet final
states are produced copiously and complicate the search for new physics.

Tree amplitudes can be used to predict cross sections at leading order (LO)
in the perturbative expansion in the QCD coupling $\alpha_s$.
Such results are already available numerically for a wide
variety of processes.  Programs such as
{\sc MadGraph}~\cite{Stelzer:1994ta,Alwall:2007st},
{\sc CompHEP}~\cite{Pukhov:1999gg},
and {\sc AMEGIC++\/}~\cite{Krauss:2001iv} are based on fast
numerical evaluation of Feynman diagrams.  Other methods include
the Berends-Giele (off shell) recursion
relations~\cite{Berends:1987me}, as implemented for example in
{\sc COMIX}~\cite{Gleisberg:2008fv}, and the related
{\sc ALPHA}~\cite{Caravaglios:1995cd,Caravaglios:1998yr}
and {\sc HELAC}~\cite{Kanaki:2000ey,Cafarella:2007pc} algorithms
based on Dyson-Schwinger equations, as well as
{\sc O'Mega/WHIZARD}~\cite{Moretti:2001zz,Kilian:2007gr}.  The computation
time required in these latter methods scales quite well with the number of legs.

On the formal side, the properties of scattering amplitudes have long
provided numerous clues to hidden symmetries and dynamical structures in gauge
theory.  It was recognized early on that tree amplitudes in gauge theory
are effectively supersymmetric~\cite{Parke:1985pn,Kunszt:1985mg}, so
that they obey supersymmetric $S$-matrix Ward
identities~\cite{Grisaru:1976vm,Grisaru:1977px}.
Soon thereafter, Parke and Taylor~\cite{Parke:1986gb} discovered a
remarkably simple formula for the 
maximally-helicity-violating (MHV) amplitudes for $n$-gluon scattering,
which was proven by Berends and Giele~\cite{Berends:1987me},
and soon generalized to ${\cal N}=4$ super Yang-Mills theory (SYM) by
Nair~\cite{Nair:1988bq}.

Later, it was found that this simplicity also extends to the loop level,
at least for ${\cal N}=4$ super Yang-Mills
theory~\cite{Bern:1994zx,Bern:1994cg}.  These results were obtained
using the unitarity method, which constructs loop amplitudes by sewing
together tree amplitudes (for recent reviews see
refs.~\cite{Bern:2007dw,Berger:2009zb}).   After Witten~\cite{Witten:2003nn}
reformulated gauge theory in terms of a topological string propagating in
twistor space, there was a huge resurgence of interest in uncovering
new properties of scattering amplitudes and developing new
methods for their efficient computation.  Among other developments,
Britto, Cachazo, Feng and Witten proved a new type of recursion
relation~\cite{Britto:2004ap,Britto:2005fq} for gauge theory.
In contrast to the earlier off-shell recursion relations, the BCFW
relation uses only on-shell lower-point amplitudes, evaluated at complex
momenta.  A particular solution to this recursion relation was found for
an arbitary number of gluons in the split-helicity configuration
$({-}\cdots{-}{+}\cdots{+})$~\cite{Britto:2005dg}.

The BCFW recursion relation was then recast as a super-recursion
relation for the tree amplitudes of ${\cal N}=4$ super Yang-Mills
theory, which involves shifts of Grassmann parameters as well as
momenta~\cite{ArkaniHamed:2009dn}. A related construction is given in
ref.~\cite{Bianchi:2008pu}.  The super-recursion relation of
ref.~\cite{ArkaniHamed:2009dn} was solved for arbitrary external
states by Drummond and one of the present
authors~\cite{Drummond:2008cr}.  Tree-level super-amplitudes have a
dual superconformal
invariance~\cite{Drummond:2008vq,Brandhuber:2008pf}, and the explicit
solution does indeed have this symmetry~\cite{Drummond:2008cr}.
It is written in terms of dual superconformal invariants, which are a
straightforward generalization of those that first appeared in
next-to-MHV (NMHV)
super-amplitudes~\cite{Drummond:2008vq,Drummond:2008bq}.
This dual superconformal invariance
of tree-level amplitudes is a hallmark of the integrability of
planar $\mathcal{N}=4$ SYM, as it closes with the standard superconformal
symmetry into an infinite-dimensional symmetry of Yangian
type~\cite{Drummond:2009fd} (a recent review is ref.~\cite{Drummond:2010ep}).

The purpose of this paper is to illustrate how these more recent formal
developments can reap benefits for phenomenological applications in QCD.
In particular, we will evaluate the solution in ref.~\cite{Drummond:2008cr}
by carrying out the integrations over Grassmann parameters that are
needed to select particular external states.  In addition, we will show how
to extract tree-level QCD amplitudes from the amplitudes of
${\cal N}=4$ super Yang-Mills theory.  While this extraction is simple for
pure-gluon amplitudes, and those with a single massless quark line, it becomes
a bit more intricate for amplitudes with multiple quark lines of different
flavors, because of the need to forbid the exchange of scalar particles,
which are present in ${\cal N}=4$ super Yang-Mills theory but not in QCD.

Although, as mentioned above, there are currently many numerical programs
available for computing tree amplitudes efficiently, the existence of
analytic expressions may provide a yet more efficient approach in some
contexts.  In fact, the formulae provided in this paper have already
served a practical purpose:  They were used to evaluate contributions
from real emission in the NLO corrections to the cross section for
producing a $W$ boson in association with four jets at the
LHC~\cite{Berger:2010zx}.  This process forms an important background to
searches for various kinds of new physics, including supersymmetry.
The real-emission corrections require evaluating nine-point tree amplitudes
at a large number of different phase-space points (on the order of $10^8$),
in order to get good statistical accuracy for the Monte Carlo integration
over phase space.

In principle, QCD tree amplitudes can also be used to speed up the evaluation
of one-loop amplitudes, when the latter are constructed from tree amplitudes
using a numerical implementation of generalized unitarity.  Many different
generalized unitarity cuts, and hence many different tree amplitudes,
are involved in the construction of a single one-loop amplitude.
The tree amplitudes described here enter directly into the construction
of the ``cut-constructible'' part~\cite{Bern:1994cg} of one-loop amplitudes
in current programs such as
{\sc CutTools}~\cite{Ossola:2006us,Ossola:2007ax,Ossola:2008xq},
{\sc Rocket}~\cite{Ellis:2007br,Giele:2008ve,Giele:2008bc} and
{\sc BlackHat}~\cite{Berger:2008sj}.
On the other hand, the computation-time bottleneck in these programs
often comes from the so-called ``rational'' terms.  When these terms are
computed using only unitarity, it is via unitarity in $D$
dimensions~\cite{Bern:1995db,Bern:1996ja,Anastasiou:2006jv,Britto:2007tt,%
Giele:2008ve,Giele:2008bc,Badger:2008cm},
not four dimensions.
The amplitudes presented here are four-dimensional ones, so they cannot be
used directly to alleviate this bottleneck for the $D$-dimensional unitarity
method.  However, in the numerical
implementation of loop-level on-shell recursion
relations~\cite{Bern:2005cq,Berger:2006ci} for the
rational part in {\sc BlackHat}~\cite{Berger:2008sj},
or in the OPP method used in
{\sc CutTools}~\cite{Ossola:2006us,Ossola:2007ax,Ossola:2008xq}, 
there are no $D$-dimensional trees, so this is not an issue.

An interesting avenue for future research would be to try to generalize the
results presented in this paper to QCD amplitudes containing massive quarks,
or other massive colored states.  Massive quark amplitudes are of interest
because, for example, processes that produce top quarks in association
with additional jets can form important backgrounds to new physics at the LHC.
States in $\mathcal{N}=4$ SYM can be given masses through a super-Higgs
mechanism.  This mechanism was explored recently in the context of infrared
regulation of $\mathcal{N}=4$ SYM loop amplitudes~\cite{Alday:2009zm}.
However, it should be possible to generate the appropriate tree
amplitudes with massive quarks, or other massive states, from the same kind
of setup, once one solves the appropriate super-BCFW recursion relations.

The remainder of this paper is organized as follows. After introducing the
standard technology of color-ordered amplitudes and spinor helicity we explain
the strategies of how to extract QCD tree amplitudes with massless quarks 
from $\mathcal{N}=4$ SYM in section 3. We also discuss how to convert
these amplitudes into trees with one electroweak vector boson.
Sections 4 through 6 are devoted to
stating the general analytical formulae for gluon-gluino $n$-parton amplitudes
in $\mathcal{N}=4$ SYM, which are proven in section 7. In appendix A 
we provide a collection of explicit results for pure-gluon trees. 
Explicit formulae for trees involving up to six fermions are displayed in 
appendix B. Finally, appendix C is
devoted to a documentation of our {\tt Mathematica} package {\tt GGT} which 
implements all of the results of this paper and yields the analytical 
expressions for an arbitrary flavored gluon-gluino tree amplitude
in $\mathcal{N}=4$ SYM.  The package is included in
the {\tt arXiv.org} submission of this article and may also be
downloaded from {\tt http://qft.physik.hu-berlin.de}.

\section{Color-ordering and spinor-helicity formalism}
\label{ColorSpinorSection}

Tree-level gluon amplitudes in non-abelian gauge theories may be conveniently
separated into a sum of terms, each composed of a simple prefactor
containing the color indices, multiplied by a kinematical factor
known as a partial or color-ordered amplitude.  For an $n$-gluon
amplitude one has
\begin{equation}
{\cal A}^{\text{tree}}_{n}(\{p_{i},h_{i},a_{i}\})
= g^{n-2}\sum_{\sigma\in S_{n}/Z_{n}}
\Tr(T^{a_{\sigma(1)}}\ldots T^{a_{\sigma(n)}})\, 
A_{n}(\sigma(1)^{h_{\sigma(1)}}\ldots \sigma(n)^{h_{\sigma(n)}} )\, ,
\label{colorordering}
\end{equation}
with the argument $i^{h_{i}}$ of the partial amplitude $A_{n}$ 
denoting an outgoing gluon of light-like momentum $p_{i}$ and helicity 
$h_{i}=\pm 1$, $i\in[1,n]$. The $su(N_c)$ generator matrices
$T^{a_i}$ are in the fundamental representation, and are normalized so
that ${\rm Tr}(T^a T^b) = \delta^{ab}$.

Color-ordered amplitudes of massless particles are most compactly expressed 
in the spinor-helicity formalism. Here all four-momenta are written as 
bi-spinors via
\begin{equation}
\pslash^{\alpha\dot\alpha}= \sigma_{\mu}^{\alpha\dot\alpha}\, p^{\mu},
\end{equation}
where we take $\sigma^{\mu}=(\mathbf{1},\vec{\sigma})$ with
$\vec{\sigma}$ being the $2\times 2$ Pauli spin matrices. Light-like
vectors are then expressed via the product of two spinors
\begin{equation}
\pslash^{\alpha\dot\alpha}=\lambda^{\alpha}\, \tilde \lambda^{\dot\alpha}\, .
\end{equation}
For real momenta with Lorentz signature we have
$\tilde \lambda=\pm \lambda^{\ast}$, with the sign being determined
by the energy component of $p$. For complex momenta the spinors $\lambda$ and
$\tilde\lambda$ are independent. Our convention is such that all
gluons are outgoing. Then in \eqn{colorordering} each
color-ordered leg is specified by a choice of spinors $\lambda_{i}$
and $\tilde\lambda_{i}$ along with a helicity $h_{i}=\pm 1$. Given
this data the associated polarization vectors may be reconstructed
from the expressions
\begin{equation}
\epsilon_{+, \, i}^{\alpha\dot\alpha}
 = \frac{\tilde\lambda_{i}^{\dot\alpha}\, \mu_{i}^\alpha}
{\vev{\lambda_{i}\mu_{i}}}\, , \qquad
\epsilon_{-, \, i}^{\alpha\dot\alpha}
 = \frac{\lambda_{i}^{\alpha}\, \tilde\mu_{i}^{\dot\alpha}}
{\bev{\lambda_{i}\mu_{i}}}\, ,
\end{equation}
where $\mu_{i}^{\alpha}\tilde\mu_{i}^{\dot\alpha}$ are auxiliary
momenta and we use the standard notation
$\vev{\lambda\mu}=\epsilon_{\alpha\beta}\lambda^{\alpha}\mu^{\beta}$
and
$\bev{\lambda\mu}
=\epsilon_{\dot\alpha\dot\beta}\lambda^{\dot\alpha}\mu^{\dot\beta}$.
Moreover we shall often use the abbreviated forms
$\vev{ij}=\vev{\lambda_{i}\lambda_{j}}$ and
$\bev{ij}=\bev{\tilde\lambda_{i}
\tilde\lambda_{j}}$ in the sequel.
As an essential building block of the general tree-level scattering
formula we introduce the dual coordinates or region momenta
$x_{ij}^{\alpha\dot\alpha}$ via
\begin{equation}
x_{ij}^{\alpha \dot{\alpha}}
\ :=\ ( \pslash_{i}+ \pslash_{i+1} + \cdots
 + \pslash_{j-1} )^{\alpha \dot{\alpha}}
= \sum_{k=i}^{j-1}\, \lambda_{k}^{\alpha} \tilde{\lambda}_{k}^{\dot\alpha}
 \, , \qquad 
\, \qquad i<j \, ,
\end{equation}
$x_{ii} = 0$, and $x_{ij} = -x_{ji}$ for $i>j$.
We then define the scalar quantities
\begin{equation}
\langle n a_{1} a_{2} \ldots a_{k} |a\rangle := 
\langle n| x_{na_{1}} x_{a_{1}a_{2}}\ldots x_{a_{k-1}a_{k}}|a\rangle \, ,
\label{defvev}
\end{equation}
which we will use frequently in the following.  In fact all amplitudes
can be expressed in terms of the quantities
$\langle n a_{1} a_{2} \ldots a_{k} |a\rangle$
and the spinor products $\ang{i}{j}$.

As an example of the notation and in order to give a flavor of the 
kinds of results we obtain, we present a compact formula for the 
$n$-point NMHV pure gluon amplitude in QCD
\begin{multline}\label{intro-nmhv-example}
A_n^{\mathrm{NMHV}}(i_{1},i_{2},n)
=\frac{\delta^{(4)}(p)}{\ang{1}{2}\dots \ang{n}{1}}\times \\
\begin{aligned}
 &  \Bigg[  \sum_{i_{1} < s \leq i_{2} < t\leq n-1}\tilde{R}_{n;st}
   \Bigl(   \ang{n}{i_{1}} \vev{nts|i_{2}}  \Bigr)^{4}
        +    \sum_{i_{1} < s <   t \leq i_{2}}\tilde{R}_{n;st}
   \Bigl( \ang{i_{2}}{n} \ang{n}{i_{1}} x_{st}^2 \Bigr)^{4}  \\
     &+\sum_{2\leq  s \leq i_{1} < i_{2} <   t  \leq n-1}\tilde{R}_{n;st}
   \Bigl( \ang{ i_{2}}{i_{1}} \vev{n ts|n} \Bigr)^{4} 
         + \sum_{2\leq  s \leq i_{1}   <   t \leq i_{2}}\tilde{R}_{n;st}
   \Bigl(  \ang{n}{i_{2}} \vev{nst|i_{1}} \Bigr)^{4} 
        \Bigg]  \,.
\end{aligned}
\end{multline}
Here $i_{1}, i_{2}$ and $n$ correspond to the positions of the three
negative-helicity gluons. (Using cyclic symmetry, we have put one of them
at position $n$ without loss of generality.)  The quantities
$\tilde{R}_{n;st}$ are simply given by 
\begin{equation}
\tilde R_{n; st}:= \frac{1}{x_{st}^{2}}\, 
\frac{\vev{s(s-1)}}{\langle n t s |s\rangle\,
\langle n t s |s-1\rangle}
\frac{\vev{t(t-1)}}{\langle n s t |t\rangle \,
\langle n s t |t-1\rangle} \, .
\label{Rtilde1}
\end{equation}
with $\tilde R_{n; st} := 0$ for $t=s+1$ or $s=t+1$.  Note that the above
formula is given for an arbitrary number of gluons $n$. In realistic
cases this number is usually small, say of the order of $9$, in which
case relatively few terms are produced by the nested sums in 
\eqn{intro-nmhv-example}.


\section{From \texorpdfstring{$\mathcal{N}=4$}{N=4} SYM to QCD tree amplitudes}
\label{sect-SYMtoQCD}

In this section we discuss how to assign quantum numbers for external
states in $\mathcal{N}=4$ SYM in order to generate tree amplitudes for
QCD with massless quarks.  We then discuss the generation of tree
amplitudes including an electroweak vector boson ($W$, $Z$ or virtual
photon).  From the point of view of tree amplitudes, there are two
principal differences between $\mathcal{N}=4$ SYM and massless QCD.
First of all, the fermions in $\mathcal{N}=4$ SYM, the gluinos,
are in the adjoint representation of $su(N_c)$, rather than the 
fundamental representation, and come in four flavors.
Secondly, the $\mathcal{N}=4$ SYM theory contains six massless scalars in the
adjoint representation.

Because we use color-ordered amplitudes, as discussed in
\secref{ColorSpinorSection}, the first difference is fairly unimportant.
Quark amplitudes can be assembled from the same
color-ordered amplitudes as gluino amplitudes, weighted with
different color factors.  For example, the color decomposition for
amplitudes with a single quark-anti-quark pair, and the remaining $(n-2)$
partons gluons is,
\begin{equation}
{\cal A}^{\text{tree}}_{n}(1_{\bar{q}},2_q,3,\ldots,n)
 \ =\  g^{n-2} \sum_{\sigma\in S_{n-2}}
   (T^{a_{\sigma(3)}}\ldots T^{a_{\sigma(n)}})_{i_2}^{~\ib_1}\
    A_n^\tree(1_{\bar{q}},2_q,\sigma(3),\ldots,\sigma(n))\,.
\label{qqgluecolorordering}
\end{equation}
The color-ordered amplitudes appearing in \eqn{qqgluecolorordering} are
just the subset of two-gluino-$(n-2)$-gluon amplitudes in which the two
gluinos are adjacent.

Amplitudes with more quark-anti-quark pairs have a somewhat
more intricate color structure involving multiple strings of $T^a$
matrices, as explained in ref.~\cite{Mangano:1990by}.
In this case, some of the color factors also include explicit factors of 
$1/N_c$, as required to project out the $su(N_c)$-singlet state for
gluon exchange between two different quark lines.  However, all of the
required kinematical coefficients can still be constructed from
suitable linear combinations of the color-ordered amplitudes for
$2k$ external gluinos and $(n-2k)$ external gluons.

The main goal of this section will be to
illustrate how to choose the flavors of the external gluinos in order
to accomplish two things: (1) avoid the internal exchange of massless
scalars, and (2) allow all fermion lines present to be for distinct
flavors.  (In some cases one may want amplitudes with (partially) identical
fermions; these can always be constructed from the distinct-flavor
case by summing over the relevant exchange-terms, although it may be
more efficient to compute the identical-fermion case directly.)  We
will accomplish this goal for amplitudes containing up to
four separate fermion lines, that is, eight external fermion states.

In gauge theory, tree amplitudes that contain only external gluons
are independent of the matter states in the
theory~\cite{Parke:1985pn,Kunszt:1985mg}; hence they
are identical between $\mathcal{N}=4$ SYM and QCD.
The reason is simply that the vertices that couple gluons to
the other states in the theory always produce the fermions and scalars
in pairs. There are no vertices that can destroy all the fermions and
scalars, once they have been produced.  If a fermion or scalar
were produced at any point in a tree diagram, it would have to
emerge from the diagram, which would no longer have only external
gluons.  In other words, the pure-gluon theory forms a closed
subsector of $\mathcal{N}=4$ SYM.

Another closed subsector of $\mathcal{N}=4$ SYM is $\mathcal{N}=1$
SYM, which contains a gluon and a single gluino.  Let $g$ denote the
gluon, $\tilde{g}_A$, $A=1,2,3,4$, denote the four gluinos, and
$\phi_{AB}=-\phi_{BA}$ denote the six real scalars of $\mathcal{N}=4$
SYM.  Then the $\mathcal{N}=1$ SYM subsector is formed by
$(g,\tilde{g}_1)$.  The reason it is closed is similar to the
pure-gluon case just discussed: There are vertices that produce states
other than $(g,\tilde{g}_1)$, but they always do so in pairs.  For
example, the Yukawa coupling $\phi^{AB} \tilde{g}_A \tilde{g}_B$,
$A\neq B$, can convert $\tilde{g}_1$ into a scalar and a gluino each
carrying an index $B\neq1$.  However, this index cannot be destroyed
by further interactions.

\begin{figure}[t]
 \centering
 \raisebox{0.5cm}{\includegraphics[width=9cm]{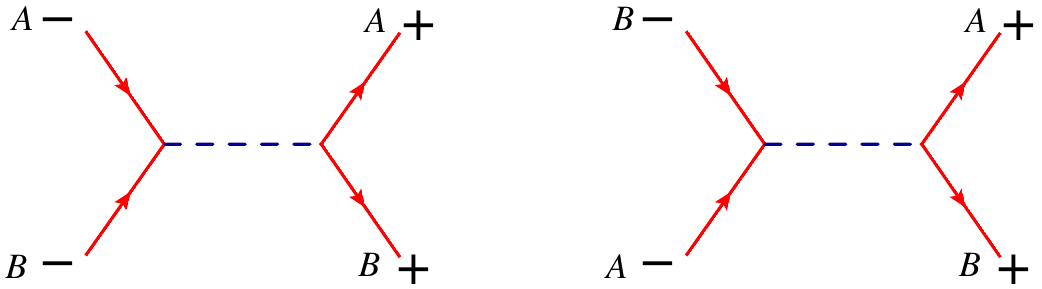}}
 \caption{Unwanted scalar exchange between fermions of different flavors,
$A\neq B$.}
\label{fig-fourqscalar}
\end{figure}

The fact that $\mathcal{N}=1$ SYM forms a closed subsector of
$\mathcal{N}=4$ SYM, in addition to color ordering,
immediately implies that any
color-ordered QCD tree amplitude for gluons, plus arbitrarily many
quarks of a single flavor, is given directly by the corresponding
amplitude (with $\tilde{g}_1$ replacing the single quark flavor)
evaluated in $\mathcal{N}=4$ SYM.  The less trivial QCD amplitudes to
extract are those for multiple fermion flavors, primarily because of
the potential for intermediate scalar exchange induced by the Yukawa
coupling $\phi^{AB} \tilde{g}_A \tilde{g}_B$.  Figure~\ref{fig-fourqscalar}
illustrates scalar exchange in an amplitude with four fermions belonging 
to two different flavor lines, $A \neq B$.

\begin{figure}[t]
 \centering
 \raisebox{0.5cm}{\includegraphics[width=12cm]{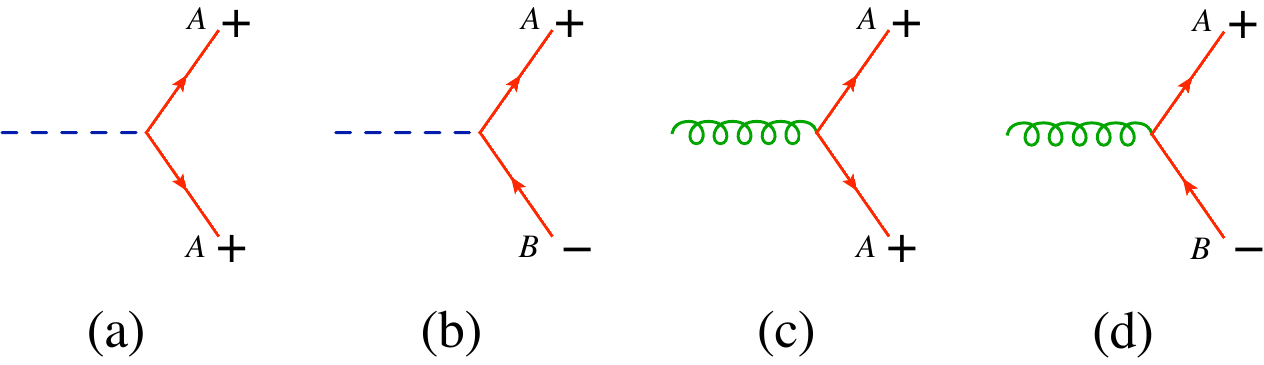}}
 \caption{These vertices all vanish, as explained in the text.
This fact allows us to avoid scalar exchange and control the flow of
fermion flavor.}
\label{fig-vanishing-vertices}
\end{figure}

The key to avoiding such unwanted scalar exchange is provided by
figure~\ref{fig-vanishing-vertices}, which shows four types of
vertices that could potentially couple fermion pairs to scalars and
gluons.  However, all four types of vertices vanish.  (Recall that
helicities are labeled in an all-outgoing convention.) Case (a)
vanishes because the Yukawa interaction only couples gluinos of
different flavors, $A\neq B$.  Cases (b) and (c) vanish because of
fermion helicity conservation for the gauge interactions, and a
helicity flip for the Yukawa coupling.  Case (d) vanishes because
gluon interactions do not change flavor.

\begin{figure}[t]
 \centering
 \raisebox{0.5cm}{\includegraphics[width=15cm]{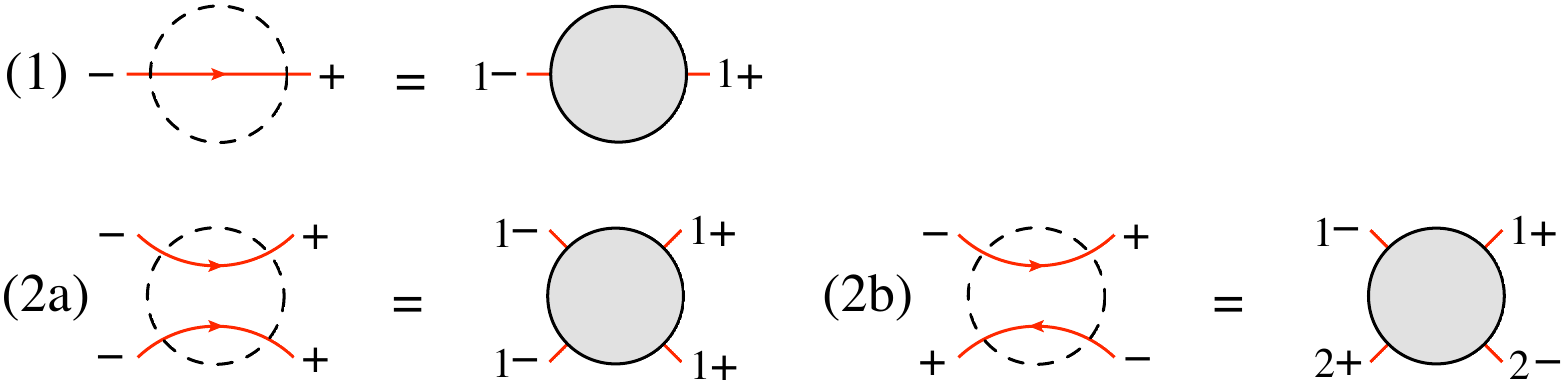}}
 \caption{The possible fermion-line configurations for amplitudes
with one fermion line, case (1), or two fermion lines, cases (2a) and (2b).}
\label{fig-fermions12}
\end{figure}

Because the emission of gluons from fermions does not change their
helicity or flavor, in analyzing whether scalar exchange can be
avoided, as well as the pattern of fermion flavor flow, one can ignore
the gluons altogether.  For example, figure~\ref{fig-fermions12} shows
the possible cases for amplitudes with one or two fermion lines.  The
left-hand side of the equality shows the desired (color-ordered)
fermion-line flow and helicity assignment for a QCD tree amplitude.
All gluons have been omitted, and all fermion lines on the left-hand
side are assumed to have distinct flavors.  The right-hand side of the
equality displays a choice of gluino flavor that leads to the desired
amplitude.  All other one- and two-fermion-line cases are related to
the ones shown by parity or cyclic or reflection symmetries.

The one-fermion line, case (1), is trivial because $\mathcal{N}=1$ SYM
forms a closed subsector of $\mathcal{N}=4$ SYM.  In case (2a) we must
choose all gluinos to have the same flavor; otherwise a scalar would
be exchanged in the horizontal direction.  Here, helicity conservation
prevents the exchange of an unwanted gluon in this direction, keeping
the two flavors distinct as desired.  In case (2b), we must use two
different gluino flavors, as shown; otherwise helicity conservation
would allow gluon exchange in the wrong channel, corresponding to
identical rather than distinct quarks.

More generally, in order to avoid scalar exchange, if two
color-adjacent gluinos have the same helicity, then we should choose
them to have the same flavor.  In other words, we should forbid all
configurations of the form $(\ldots,A^+,B^+,\ldots)$ and
$(\ldots,A^-,B^-,\ldots)$ for $A\neq B$, where $A^\pm$ stands for the
gluino state $\tilde{g}_A^\pm$.  While this is necessary, it is not
sufficient.  For example, we also need to forbid configurations such
as $(\ldots,A^+,C^\pm,C^\mp,B^+,\ldots)$, because the pair
$(C^\pm,C^\mp)$ could be produced by a gluon splitting into this pair,
which also connects to the $(A^+,B^+)$ fermion line.  As a secondary
consideration, if two color-adjacent gluinos have opposite helicity,
then we should choose them to have the same flavor or different flavor
according to the desired quark flavor flow on the left-hand side.
However, it will not always be possible to choose them to have a
different flavor.

\begin{figure}[t]
 \centering
 \raisebox{0.5cm}{\includegraphics[width=15cm]{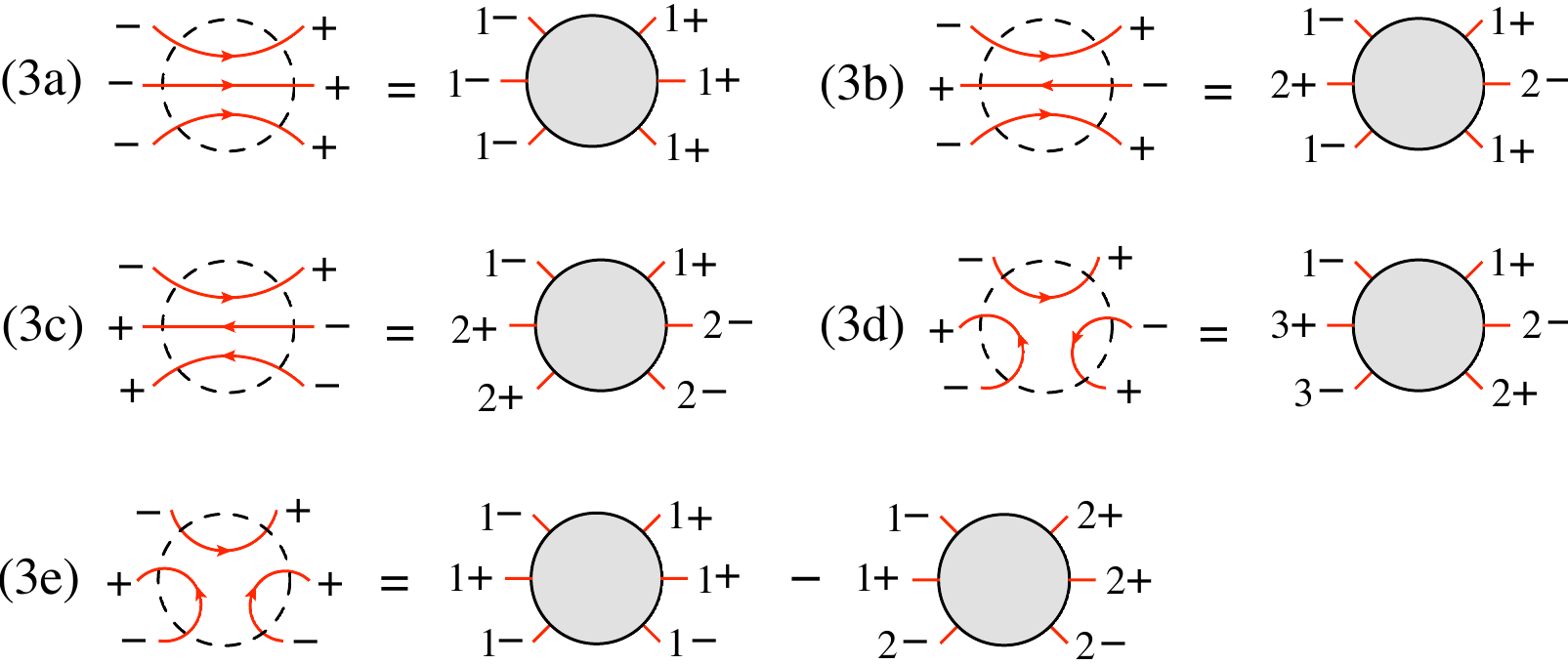}}
 \caption{The possible fermion-line configurations for amplitudes
with three fermion lines.}
\label{fig-fermions3}
\end{figure}

With these properties in mind, we can now examine
figure~\ref{fig-fermions3}, which shows solutions for the
three-fermion-line cases.  (Again, all other three-fermion-line cases
are related to the ones shown by parity or cyclic or reflection
symmetries.)  Using the properties of the vanishing vertices in
figure~\ref{fig-vanishing-vertices}, it is straightforward to show that
only the desired QCD tree amplitudes on the left are produced by the
gluino assignments on the right.  The most subtle case is (3e).  The
two pairs of adjacent identical-helicity quarks force all the gluino
flavors to be the same.  However, this choice does not result in three
distinct fermion lines in the pattern desired.  There is a ``wrong''
fermion-line configuration which, fortunately, coincides with a
permutation of case (3c).  Hence we can subtract off this solution, as
the second term on the right-hand side of (3e).

\begin{figure}[t]
 \centering
 \raisebox{0.5cm}{\includegraphics[width=17cm]{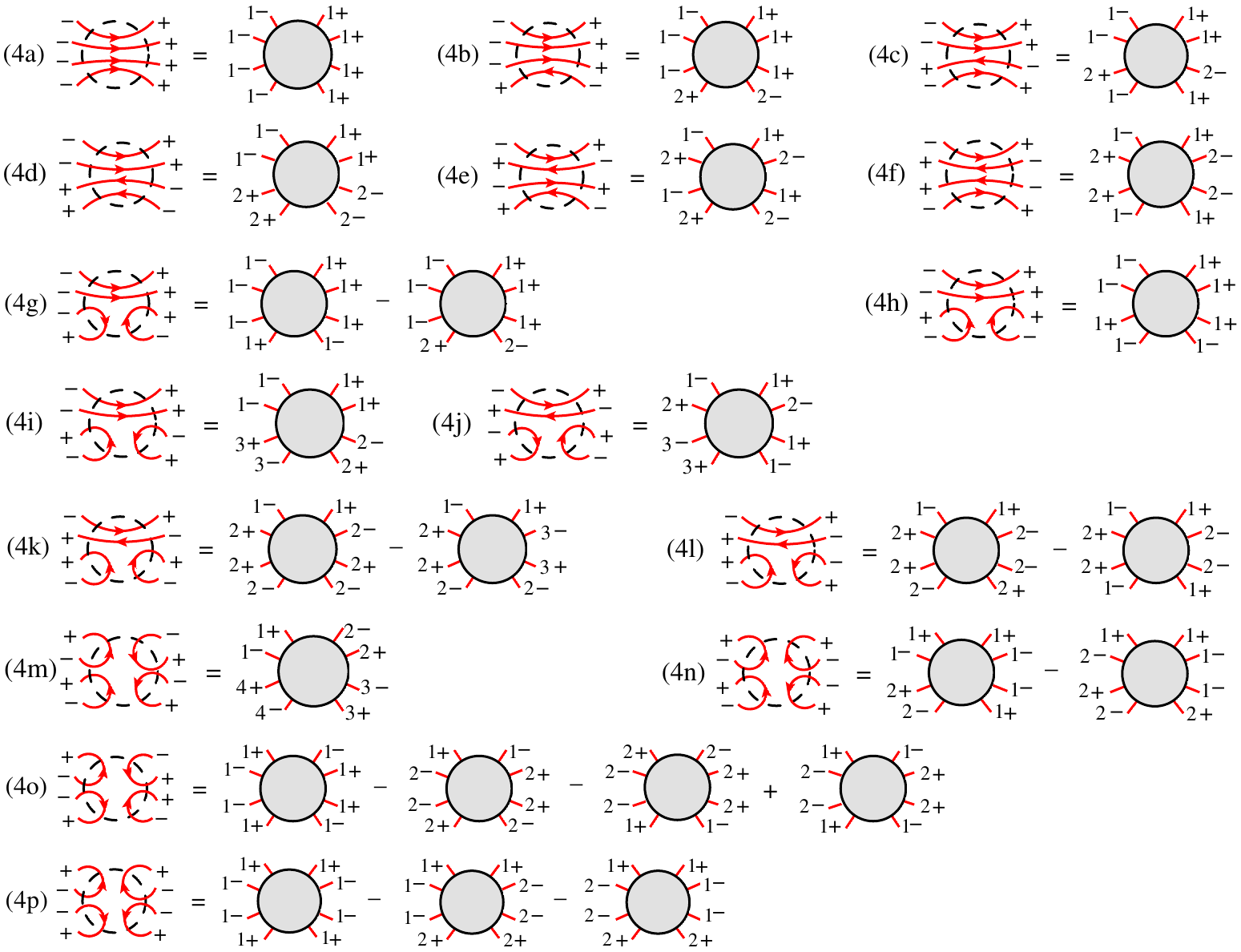}}
 \caption{The possible fermion-line configurations for amplitudes
with four fermion lines.}
\label{fig-fermions4}
\end{figure}

We have also examined the cases with four different fermion lines.
There are 16 cases, shown in figure~\ref{fig-fermions4}, and each has
a solution. Four of the cases, (4g), (4k), (4l) and (4n), require two
different terms, analogous to case (3e).  Case (4p) requires three
terms to remove all identical fermions; and case (4o) requires four
separate terms.  Case (4o) can be written as
\begin{align}
&A^{\rm QCD}
(\bar{q}_1^-,q_1^+,\bar{q}_2^-,q_2^+,q_3^+,\bar{q}_3^-,q_4^+,\bar{q}_4^-)
= A^{\mathcal{N}=4\ SYM}(1^-,1^+,1^-,1^+,1^+,1^-,1^+,1^-) \nonumber\\
&\=\null -[ A^{\mathcal{N}=4\ SYM}(2^-,1^+,1^-,2^+,2^+,2^-,2^+,2^-)
  - A^{\mathcal{N}=4\ SYM}(2^-,1^+,1^-,2^+,2^+,1^-,1^+,2^-) ] \nonumber\\
&\=\null - [ A^{\mathcal{N}=4\ SYM}(2^-,2^+,2^-,2^+,2^+,1^-,1^+,2^-)
  - A^{\mathcal{N}=4\ SYM}(2^-,1^+,1^-,2^+,2^+,1^-,1^+,2^-) ]   \nonumber\\
&\=\null - A^{\mathcal{N}=4\ SYM}(2^-,1^+,1^-,2^+,2^+,1^-,1^+,2^-)
   \label{worst4fcase}\\
 &= A^{\mathcal{N}=4\ SYM}(1^-,1^+,1^-,1^+,1^+,1^-,1^+,1^-)
   - A^{\mathcal{N}=4\ SYM}(2^-,1^+,1^-,2^+,2^+,2^-,2^+,2^-) \nonumber\\
&\=\null
  - A^{\mathcal{N}=4\ SYM}(2^-,2^+,2^-,2^+,2^+,1^-,1^+,2^-)
  + A^{\mathcal{N}=4\ SYM}(2^-,1^+,1^-,2^+,2^+,1^-,1^+,2^-)
  . \label{worst4fcasealt}
\end{align}
The first form of this equation indicates that three different
wrong-fermion-line configurations have to be removed.  However, all removals
can be accomplished with the help of different permutations of two other
cases, (4f) and (4l) respectively,
\begin{align}
&A^{\rm QCD}
(\bar{q}_1^-,q_1^+,\bar{q}_2^-,\bar{q}_3^-,q_4^+,\bar{q}_4^-,q_3^+,q_2^+)
 = A^{\mathcal{N}=4\ SYM}(1^-,1^+,2^-,2^-,1^+,1^-,2^+,2^+) \, ,
 \label{another4fcase}\\[+0.2cm]
&A^{\rm QCD}
(\bar{q}_1^-,q_1^+,\bar{q}_2^-,\bar{q}_3^-,q_3^+,\bar{q}_4^-,q_4^+,q_2^+)
\nonumber\\
&\quad= A^{\mathcal{N}=4\ SYM}(1^-,1^+,2^-,2^-,2^+,2^-,2^+,2^+) 
      - A^{\mathcal{N}=4\ SYM}(1^-,1^+,2^-,2^-,1^+,1^-,2^+,2^+).
\label{yetanother4fcase}
\end{align}
Case (4l) itself requires a wrong-fermion-line subtraction.

We have not yet ascertained whether any QCD tree amplitudes with more
than four fermion lines are impossible to extract from $\mathcal{N}=4$
super Yang-Mills theory.
Fortunately, for a fixed number of external partons, as one increases
the number of fermion lines the number of Feynman diagrams decreases.
Also, amplitudes with many external quarks typically contribute much
less to a multi-jet cross section than do amplitudes with more gluons
and fewer quarks.

Finally, we remark on the conversion of pure-QCD tree amplitudes, that is
amplitudes for quarks and gluons, into amplitudes that
contain in addition a single electroweak vector boson, namely a $W$, $Z$ or
virtual photon.  It is sufficient to compute the amplitude including the
decay of the boson to a fermion-anti-fermion pair.

Consider first the case
of a virtual photon which is emitted from a quark $q$ and decays to a 
charged lepton (Drell-Yan) pair $\ell^+\ell^-$.  We can extract this 
amplitude from a color-ordered amplitude with four consecutive fermions,
as follows:
\begin{equation}
A^{\gamma^*}(\ldots,q^+,\ell^-,\ell^+,\bar{q}^-,\ldots)
= A^{\rm QCD}(\ldots,q_1^+,\bar{q}_2^-,q_2^+,\bar{q}_1^-,\ldots) \, .
\label{gammastar}
\end{equation}
The color-ordering prevents gluons from being emitted from the
quark line $q_2$, or from the virtual gluon connecting $q_1$ and $q_2$.
Hence this virtual gluon is functionally identical to a virtual photon.
The only other modification comes when dressing $A^{\gamma^*}$ with
couplings.  There is a relative factor of $2(-Q^q)e^2/g^2$ when doing so,
where the ``$2$'' is related to the standard normalizations of the QED
interaction with coupling $e$, versus the QCD interaction with coupling $g$,
and $Q^q$ is the electric charge of the quark $q$.
(The lepton has charge $-1$.)

It is possible to extract the amplitude~\eqref{gammastar} a second way,
using one quark flavor instead of two,
\begin{equation}
A^{\gamma^*}(\ldots,q^+,\ell^-,\ell^+,\bar{q}^-,\ldots)
= - A^{\rm QCD}(\ldots,q_1^+,q_1^+,\bar{q}_1^-,\bar{q}_1^-,\ldots) \, .
\label{gammastar2}
\end{equation}
This alternative works because the color-ordered interaction for 
$g^* \to q^+ \bar{q}^-$ is antisymmetric under exchange of $q$ and $\bar{q}$,
and because the exchange of a gluon between identical-flavor quarks
in the wrong channel is prevented by helicity conservation.

If the virtual photon decays to other charged massless fermions,
{\it i.e.}~to a 
quark-anti-quark pair $q'\bar{q}'$, the only difference is of course to
use the appropriate quark charge, $-Q^q e^2 \to Q^q Q^{q'}e^2$.
Because helicity amplitudes are used, it is also trivial to convert the
virtual-photon amplitudes to ones for (parity-violating)
electroweak boson production. 
For the case of combined exchange of virtual photon and $Z$ boson, with
decay to a charged lepton pair, the electric charge factor has to be
replaced by 
\begin{equation}
2e^2 \Bigl( - Q^q
+ v_{L,R}^\ell v_{L,R}^q {\cal P}_Z(s_{\ell\bar{\ell}}) \Bigr)\, ,
\label{Zconversion}
\end{equation}
where $v_{L,R}^\ell$ are the left- and right-handed couplings of the
lepton to the $Z$ boson, $v_{L,R}^q$ are the corresponding quantities for the
quark,
\begin{equation}
{\cal P}_Z(s) = \frac{s}{s-M_Z^2+i\,\Gamma_Z\,M_Z}
\label{Zprop}
\end{equation}
is the ratio of $Z$ to $\gamma^*$ propagators, and $M_Z$ and $\Gamma_Z$
are the mass and width of the $Z$.

Whether $v_L$ or $v_R$ is to be used in \eqn{Zconversion}
depends on the helicity assignment, {\it i.e.}~on whether the $Z$ couples to
a left- or right-handed outgoing fermion (as opposed to anti-fermion);
see ref.~\cite{Bern:1997sc} for further details.  The case of a $W^\pm$
boson is even simpler, because there is no coupling to right-handed fermions
(and no interference with another boson).

The relevant MHV and NMHV amplitudes for
four external fermions and the rest gluons,
and for six external fermions and the rest gluons,
have been converted in the above manner into tree
amplitudes for $V q\bar{q} g \ldots g$ and $V q\bar{q} Q\bar{Q} g \ldots g$,
where $V$ stands for $W$, $Z$ or $\gamma^*$.
These NMHV amplitudes have been incorporated into
the {\sc BlackHat} library~\cite{Berger:2008sj} and used there in
conjunction with a numerical implementation \cite{Dinsdale:2006sq} 
of the BCFW (on-shell) recursion
relations~\cite{Britto:2004ap,Britto:2005fq} in order to obtain
amplitudes at the NNMHV level and beyond.
Including the MHV and NMHV formulae speeds up the numerical
recursive algorithm by a factor of about four, in the present implementation.
This approach was used to compute the real-radiative corrections entering the
recent evaluation of $pp \to W\,+\,4$ jets at NLO~\cite{Berger:2010zx},
in particular the tree amplitudes for $W q\bar{q}' ggggg$ and
$W q\bar{q}' Q\bar{Q}ggg$.  These amplitudes have nine external legs,
after decaying the $W$ boson to a lepton pair, so there are MHV, NMHV
and NNMHV configurations, but no further.  All seven-point configurations are
either MHV or NMHV, so at most two recursive steps were required to hit
an explicit formula (for example, in a schematic notation
$A_9 \to A_8\times A_3 \to A_7\times A_3 \times A_3$).

We remark that the tree-level color-ordered amplitudes entering 
subleading-color loop amplitudes can have a more general color ordering
from that required for purely tree-level applications.
For example, in the pure QCD amplitudes with a single $q\bar{q}$ pair,
only the color-ordered amplitudes in which the two fermions are adjacent
are needed in~\eqn{qqgluecolorordering}.  However, the subleading-color 
terms in the one-loop amplitudes for $q\bar{q}g\ldots g$ include many
cases in which the two fermions are not color-adjacent, and the tree-level
$q\bar{q}g\ldots g$ amplitudes that enter their unitarity cuts have the
same property.  These color-ordered amplitudes are all available in
$\mathcal{N}=4$ super Yang-Mills theory, of course.

Similarly, for computing subleading-color one-loop terms for
single-vector boson production processes like $V q\bar{q} g \ldots g$,
one needs tree amplitudes such as
$A^{\gamma^*}(\ldots,q^+,g,\ell^-,\ell^+,\bar{q}^-,\ldots)$,
in which the gluon $g$ is color-ordered with respect to the quark pair,
but not the lepton pair.  These amplitudes are not equal to any particular
color-ordered amplitude in $\mathcal{N}=4$ SYM, but one can generate them
by summing over appropriate color orderings.  For example, we have
\begin{multline}
A^{\gamma^*}(\ldots,q^+,g,\ell^-,\ell^+,\bar{q}^-,\ldots)
= A^{\rm QCD}(\ldots,q_1^+,g,\bar{q}_2^-,q_2^+,\bar{q}_1^-,\ldots)\\
+ A^{\rm QCD}(\ldots,q_1^+,\bar{q}_2^-,g,q_2^+,\bar{q}_1^-,\ldots)
+ A^{\rm QCD}(\ldots,q_1^+,\bar{q}_2^-,q_2^+,g,\bar{q}_1^-,\ldots) \,.
\label{gammastar_subl}
\end{multline}
The sum over the three permutations properly cancels out the unwanted
poles as $g$ becomes collinear with either $\ell^-$ or $\ell^+$.


\section{All gluon tree amplitudes}
\label{sect-gluons}

In this section we present the general expression for an $n$-gluon tree
amplitude, which we derive in section \ref{proof} from the solution of 
ref.~\cite{Drummond:2008cr} for a general
$\mathcal{N}=4$ SYM super-amplitude.

Amplitudes for $n$-gluon scattering are classified by the number of
negative-helicity gluons occurring in them. Tree-amplitudes with fewer
than two negative-helicity gluons vanish.  In our conventions the
gluon at position $n$ is always of negative helicity, which does not
present a restriction due to cyclicity of the
color-ordered amplitude. The Parke-Taylor
formula~\cite{Parke:1986gb} for a maximally-helicity-violating (MHV)
gluon amplitude, with the two negative-helicity gluons sitting at
positions $c_{0}\in[1,n-1]$ and $n$, then reads
\begin{equation}
 A_{n}(1^{+},\ldots,(c_{0}-1)^{+}, c_{0}^{-},(c_{0}+1)^{+},\ldots,
(n-1)^{+},n^{-}) := A^{\text{MHV}}_{n}(c_{0},n) =
\frac{\delta^{(4)}(p)\,
\ang{c_{0}}{n}^{4}}{\ang{1}{2}\ang{2}{3}\ldots\ang{n}{1}}\, ,
\end{equation}
with the total conserved momentum $p=\sum_{i=1}^{n}p_{i}$. 

Next-to-maximally-helicity-violating amplitudes of degree $p$
(N${}^{\text{p}}$MHV) then consist of $(p+2)$ negative-helicity gluons
embedded in $(n-p-2)$ positive-helicity states. We parametrize the
positions of the negative-helicity gluons in the ordered set
$(c_{0},c_{1},\ldots,c_{p},n)$ with $c_{i}\in[1,n-1]$.

The general N${}^{\text{p}}$MHV tree-amplitude then takes the form
\begin{equation}\boxed{
 \begin{array}{rl}
 A^{\text{N${}^{\text{p}}$MHV}}_{n}(c_{0},c_{1},\ldots,c_{p},n) &= 
 \displaystyle\frac{\delta^{(4)}(p)}{\vev{12}\vev{23}\ldots\vev{n1}}  \, \times
 \\
 \times \hspace{-0.8cm}
 \displaystyle \sum_{\text{all paths of length $p$}} &\hspace{-1.2 cm} 
   \displaystyle 1\cdot{\tilde R}_{n;a_{1}b_{1}}\cdot
 {\tilde R}^{\{L_{2}\};\{U_{2}\}}_{n;\{I_{2}\}; a_{2}b_{2}}\cdot \ldots \cdot 
 {\tilde R}^{\{L_{p}\};\{U_{p}\}}_{n;\{I_{p}\}; a_{p}b_{p}}
 \cdot \Bigl(\det \Xi_{n}^{\text{path}}(c_{0},\ldots,c_{p})\Bigr)^{4}
 \end{array}
 }
\label{masterglue}
\end{equation}
Let us now explain in turn the ingredients of this result, {\it i.e.}~the sum 
over paths, the $\tilde R$-functions, and the path-matrix
$\Xi_{n}^{\text{path}}$.

\begin{figure}[t]
 \centering
 \raisebox{0.5cm}{\includegraphics[width=12cm]{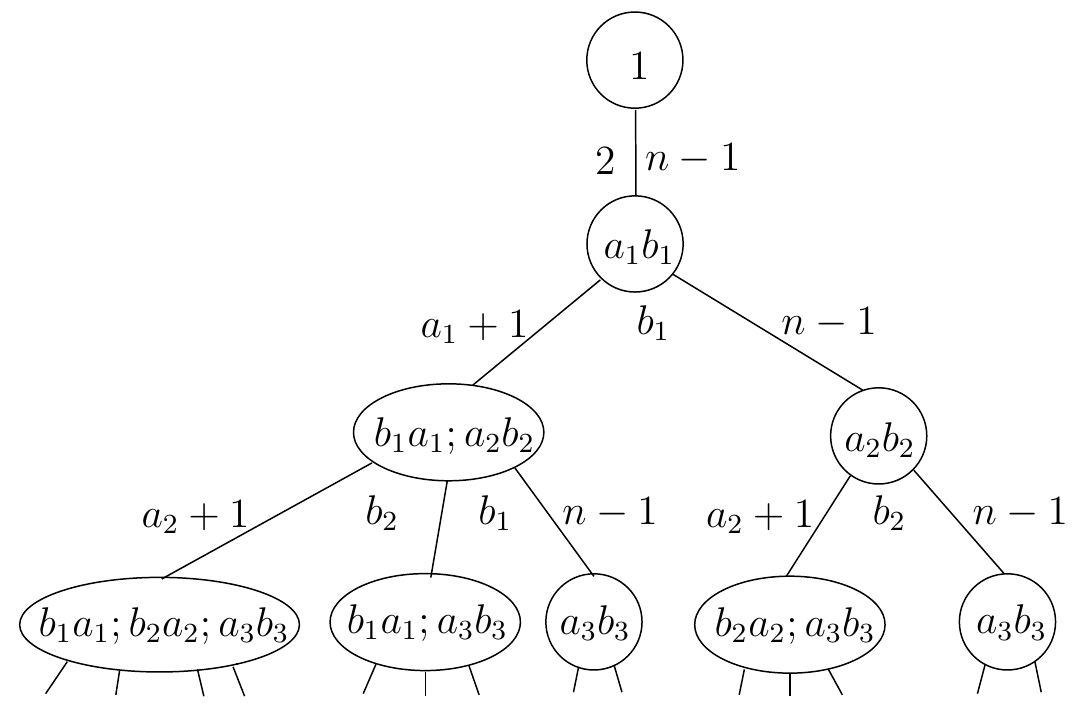}}
 \begin{tabular}{|l| l|}\hline
Length $p$ & Paths \\ \hline
0      &  $[1]$  \\ \hline
1      &  $[1]\cdot[a_{1},b_{1}]$  \\ \hline
2      &  $[1]\cdot[a_{1},b_{1}]\cdot [b_{1},a_{1};a_{2},b_{2}]$ \\
       &  $[1]\cdot[a_{1},b_{1}]\cdot [a_{2},b_{2}]$ \\ \hline
3      &  $[1]\cdot[a_{1},b_{1}]\cdot [b_{1},a_{1};a_{2},b_{2}]\cdot
 [b_{1},a_{1},b_{2},a_{2};a_{3},b_{3}]$ \\
   &  $[1]\cdot[a_{1},b_{1}]\cdot [b_{1},a_{1};a_{2},b_{2}]\cdot
 [b_{1},a_{1};a_{3},b_{3}]$ \\
  &  $[1]\cdot[a_{1},b_{1}]\cdot [b_{1},a_{1};a_{2},b_{2}]\cdot
 [a_{3},b_{3}]$ \\
  &  $[1]\cdot[a_{1},b_{1}]\cdot [a_{2},b_{2}]\cdot[b_{2},a_{2};a_{3},b_{3}]$\\
  &  $[1]\cdot[a_{1},b_{1}]\cdot [a_{2},b_{2}]\cdot[a_{3},b_{3}]$ \\ 
\hline
\end{tabular}
\caption{The rooted tree encoding the sum over paths occurring in
eq.~(\ref{masterglue}). The table shows all paths up to length 3.}
\label{rootedtree}
\end{figure}

The sum over all paths refers to the rooted tree depicted in figure
\ref{rootedtree}, introduced in ref.~\cite{Drummond:2008cr}. A path of
length 0 consists of the initial node ``1''.  A path of length $p$
leads to a sequence of $p+1$ nodes visited starting with node ``1''.
To clarify this all possible paths up to length $p=3$ are listed in
figure~\ref{rootedtree}. In general there are $(2p)!/(p!(p+1)!)$
different paths of length $p$, which is equal to the number of nodes
appearing at level $p$ in the rooted tree, since each final node
unambiguously determines a path through the rooted tree.

The $\tilde R$-functions are generalizations of \eqn{Rtilde1} and
may be written rather compactly with the help of \eqn{defvev} as
\begin{equation}
\tilde R_{n;\{I\}; ab}:=
\frac{1}{x_{ab}^{2}}\, \frac{\vev{a(a-1)}}
{\langle n\, \{I\}\, b a |a\rangle\,
\langle n \,\{I\}\, b a |a-1\rangle}
\frac{\vev{b(b-1)}}{\langle n \,\{I\}\, a b |b\rangle 
\, \langle n \,\{I\} \,a b |b-1\rangle} \, ;
\label{Rtildedef}
\end{equation}
they derive from the dual superconformal $R$-invariants introduced
in ref.~\cite{Drummond:2008cr}. In the above and in \eqn{masterglue},
$\{I\}$ is a multi-index deriving from the node in the associated path
with the last pair of indices stripped off,
{\it e.g.}~$\{I_{3}\}= \{b_{1},a_{1},b_{2},a_{2}\}$ for the last node of the
first path of length $p=3$.

In \eqn{masterglue} we used a further piece of notation, namely 
$\tilde R$-functions with upper indices, which start to appear at the NNMHV
level, and which we now define.
Generally the $\tilde R$-functions appear in the amplitude with an
ordered summation over the last pair of indices,
\be
\sum_{L\leq a<b\leq U} \tilde R_{n;\{I\};ab} \ .
\ee
$\tilde R$-functions with upper indices indicate a special behavior for the
boundary terms in this sum when $a=L$ or $b=U$. Specifically if one has
\be
\label{summedRtilde}
\sum_{L\leq a<b\leq U}
\tilde R^{l_{1},\ldots, l_{p}; u_{1},\ldots, u_{q}}_{n;\{I\};ab}\, ,
\ee
and the boundary of a summation is reached, then the occurrence of the
spinor $|a-1\rangle$ or $|b\rangle$ in the $\tilde R$-function without
upper indices (\ref{Rtildedef}) is replaced by a novel spinor depending
on the upper indices as follows
\begin{align}
\langle L-1| &\longrightarrow \langle \xi_{L}|
:= \langle n |x_{nl_{1}}x_{l_{1}l_{2}}\ldots x_{l_{p-1}l_{p}}
\qquad\quad \text{for $a=L$,}\nonumber\\
\langle U| &\longrightarrow \langle \xi_{U}|:=
\langle n |x_{nu_{1}}x_{u_{1}u_{2}}\ldots x_{u_{q-1}u_{q}}
\qquad \text{for $b=U$.}
\end{align}
Effectively this amounts to the following formula for the upper-indexed
$\tilde R$-functions of \eqn{summedRtilde},
\be
\tilde R^{l_{1},\ldots, l_{p};u_{1},\ldots, u_{q}}_{n;\{I\};ab}
=\begin{cases}
\tilde R_{n;\{I\};ab}\cdot\frac{\vev{a\, \xi_{L}}}{\vev{n\{I\}ba|\xi_{L}}}
\frac{\vev{n\{I\}ba|a-1}}
{\vev{a(a-1)}} \,  & \text{for $a=L$,} \cr
 \tilde R_{n;\{I\};ab}\cdot
\frac{\vev{\xi_{U}(b-1)}}{\vev{n\{I\}ab|\xi_{U}}}
\frac{\vev{n\{I\}ab|b}}{\vev{b(b-1)}}
 & \text{for $b=U$,} \cr
\tilde R_{n;\{I\};ab}\cdot
\frac{\vev{a\, \xi_{L}}}{\vev{n\{I\}ba|\xi_{L}}}\frac{\vev{n\{I\}ba|a-1}}
{\vev{a(a-1)}}\frac{\vev{\xi_{U}(b-1)}}{\vev{n\{I\}ab|\xi_{U}}}
\frac{\vev{n\{I\}ab|b}}{\vev{b(b-1)}} & \text{for $a=L$ and $b=U$,} \cr
\tilde R_{n;\{I\};ab} & \text{else.} \cr
\end{cases}
\ee
In particular there is a term in the double sum where both $a=L$ and $b=U$ are 
reached and both replacements are to be made.  The rules for constructing
the sets of upper indices $l_1,\ldots,l_p;u_1,\ldots,u_q$
in~\eqn{summedRtilde} from the rooted tree are given in
ref.~\cite{Drummond:2008cr}.

To write down the path-matrix $\Xi_{n}^{\text{path}}$ we furthermore
need to define the quantities
\begin{align}
(\Xi_{n})^{c_{i}}_{0} &:= \vev{n c_{i}} \,,
\\
(\Xi_{n})^{c_{i}}_{ab} &:= 
\vev{nba|c_{i}}\, \chi_{[a,b-1]}(c_{i}) -
x_{ab}^{2}\,\vev{n c_{i}}\, \chi_{[b,n-1]}(c_{i}) \,,
\\
(\Xi_{n})^{c_{i}}_{\{b_{1},a_{1},\ldots , b_{r},a_{r}\};ab}
&:= \vev{nb_{1}a_{1}\ldots  b_{r}a_{r}\, ab|c_{i}}\, \chi_{[a,b-1]}(c_{i})
- x_{ab}^{2}\, \vev{n b_{1} a_{1}\ldots  b_{r}a_{r}\,|c_{i}}\,
\chi_{[a_{r},a-1]}(c_{i}) \,, 
\end{align}
with the support functions 
\begin{equation}
\chi_{[a,b]}(i)=\begin{cases} 1 & \text{if}\quad i\in[a,b],
\\ 0 & \text{else.}\end{cases}
\end{equation}
Now to every node $[\{I\};ab]$ along a given path and to every
negative-helicity leg $c_{i}$ we associate the entry of the
path-matrix $(\Xi_{n})^{c_{i}}_{\{I\};ab}$.
The entries $(\Xi_{n})^{A}{}_{B}$ form a $(p+1)\times (p+1)$ matrix.
Explicitly one has
\begin{equation}
\label{Pathmatrix1}
\Xi_{n}^{\text{path}}(c_{0},\ldots,c_{p}) :=
\left (
\begin{array}{llll}
\vev{nc_{0}}  & \vev{nc_{1}} & \ldots & \vev{nc_{p}} \cr\cr
(\Xi_{n})^{c_{0}}_{a_{1}b_{1}} & (\Xi_{n})^{c_{1}}_{a_{1}b_{1}} 
                             & \ldots & (\Xi_{n})^{c_{p}}_{a_{1}b_{1}} \cr\cr
 (\Xi_{n})^{c_{0}}_{\{I_{2}\};a_{2}b_{2}}
    & (\Xi_{n})^{c_{1}}_{\{I_{2}\};a_{2}b_{2}}
                        & \ldots & (\Xi_{n})^{c_{p}}_{\{I_{2}\};a_{2}b_{2}} \cr
\,\,\,\, \vdots & \,\,\,\,\vdots & & \,\,\,\,\vdots \cr
 (\Xi_{n})^{c_{0}}_{\{I_{p}\};a_{p}b_{p}}
    & (\Xi_{n})^{c_{1}}_{\{I_{p}\};a_{p}b_{p}}
                      & \ldots & (\Xi_{n})^{c_{p}}_{\{I_{p}\};a_{p}b_{p}} \cr
\end{array}\right ) \ .
\end{equation}
Although $\tilde R$ and $\Xi^{\text{path}}_{n}$ look rather involved at
first sight, they are determined entirely through the external spinors
$\lambda_{i}$ and $\tilde\lambda_{i}$.

To clarify the construction principle let us write down the first three 
amplitudes in the N${}^{\text{p}}$MHV series explicitly:
\begin{align}
%
%
A^{\text{MHV}}_{n}(c_{0},n) &= 
\frac{\delta^{(4)}(p)}{\vev{12}\vev{23}\ldots\vev{n1}}
 \cdot \vev{nc_{0}}^{4} \ , \\
%
%
A^{\text{NMHV}}_{n}(c_{0},c_{1},n) &= 
\frac{\delta^{(4)}(p)}{\vev{12}\vev{23}\ldots\vev{n1}} 
\sum_{2\leq a_{1}<b_{1}\leq n-1} 
  {\tilde R}_{n;a_{1}b_{1}}\cdot
\left |
\begin{array}{llll}
\vev{nc_{0}}  & \vev{nc_{1}} \cr
(\Xi_{n})^{c_{0}}_{a_{1}b_{1}} & (\Xi_{n})^{c_{1}}_{a_{1}b_{1}}  \cr
\end{array}\right |^{4} \ , \\
%
%
A_n^{\text{N${}^2$MHV}}(c_{0},c_{1},c_{2},n)&=
\frac{\delta^{(4)}(p)}{\vev{12}\vev{23}\ldots\vev{n1}}\sum_{2\leq a_1<b_1<n}
{\tilde R}_{n;a_1 b_1}\cdot \nn \\ & \Biggl[\sum_{a_1+1\leq a_2<b_2\leq b_1}
{\tilde R}^{0;a_1 b_1}_{n;b_{1}a_{1};a_{2}b_{2}} 
\cdot\left |
\begin{array}{llll}
\vev{nc_{0}}  & \vev{nc_{1}} &  \vev{nc_{2}} \cr
(\Xi_{n})^{c_{0}}_{a_{1}b_{1}} & (\Xi_{n})^{c_{1}}_{a_{1}b_{1}} &
(\Xi_{n})^{c_{2}}_{a_{1}b_{1}} \cr
(\Xi_{n})^{c_{0}}_{b_{1},a_{1};a_{2}b_{2}}
 & (\Xi_{n})^{c_{1}}_{b_{1},a_{1};a_{2}b_{2}} &
(\Xi_{n})^{c_{2}}_{b_{1},a_{1};a_{2}b_{2}} \cr
\end{array}\right |^{4}\nn\\
&\, 
+\sum_{b_{1}\leq a_{2},b_{2}<n} {\tilde R}^{a_{1}b_{1};0}_{n;a_{2}b_{2}}
\cdot\left |
\begin{array}{llll}
\vev{nc_{0}}  & \vev{nc_{1}} &  \vev{nc_{2}} \cr
(\Xi_{n})^{c_{0}}_{a_{1}b_{1}} & (\Xi_{n})^{c_{1}}_{a_{1}b_{1}} &
(\Xi_{n})^{c_{2}}_{a_{1}b_{1}} \cr
(\Xi_{n})^{c_{0}}_{a_{2}b_{2}} & (\Xi_{n})^{c_{1}}_{a_{2}b_{2}} &
(\Xi_{n})^{c_{2}}_{a_{2}b_{2}} \cr
\end{array}\right |^{4}\, \Biggr] \ . 
\end{align}
In appendix \ref{explicit} we spell out the NMHV and N${}^{2}$MHV
amplitudes explicitly. We provide a {\tt Mathematica}
package {\tt GGT} with the {\tt arXiv.org} submission of this article, 
which expands the master formula~\eqref{masterglue} explicitly for 
any choice of $p$, positions $c_{i}$ and momenta
$\lambda_{i}\tilde\lambda_{i}$.
See appendix \ref{GGT} for documentation. The formula (\ref{masterglue})
is implemented by the function {\tt GGTgluon}.

It should be mentioned that in practice the number of terms arising
from the determinants of the path-matrix $\Xi^{\text{path}}_{n}$ is
often quite small, see {\it e.g.}~\eqn{intro-nmhv-example}.  Moreover,
for small $n$ the number of non-zero terms in the nested sums can be
relatively small.

\section{All single-flavor quark-anti-quark-gluon trees}
\label{sect:mono-flavor}

Turning to the gauge-theory amplitudes involving massless single-flavor
quark-anti-quark pairs we can write down a similarly general formula based
on paths along the rooted tree of figure~\ref{rootedtree}.
In an abuse of notation, we refer here to a helicity $+\frac12$ fermion
as a quark, and a helicity $-\frac12$ fermion as an anti-quark.
Looking at a color-ordered $n$-parton amplitude involving gluons and $k$
quark-anti-quark pairs, $g^{n-2k}(q\bar q)^{k}$, it is again classified
as a N${}^{\text{p}}$MHV amplitude by the number $(2+p-k)$ of
negative-helicity gluons.  In such a color-ordered amplitude we
furthermore consider an arbitrary ordering of the fermions. 
We then have $2+p+k$ `special' legs (negative-helicity gluon,
quark or anti-quark) in such an amplitude, whose position amongst the
$n$ legs we parametrize by the set
\begin{equation}
(c_{0},\ldots,c_{\alpha_{1}},\ldots , c_{\bar \beta_{1}}
\ldots , c_{\alpha_{k}},
\ldots, c_{\bar \beta_{k}},\ldots, c_{p+k},n)\, .
\label{speciallegs}
\end{equation}
Here $c_{i}$ denotes the position of a negative-helicity gluon,
$c_{\alpha_{j}}$ a quark and $c_{\bar \beta_{j}}$ an anti-quark
location.  Note that while the quark and anti-quark locations
$c_{\alpha_{i}}$ and $c_{\bar\beta_{i}}$ are considered as ordered
sets, {\it i.e.}~$c_{\alpha_{i}}<c_{\alpha_{j}}$ and
$c_{\bar\beta_{i}}<c_{\bar\beta_{j}}$ for $i<j$, there is no such
ordering in the total set $\{c_{\alpha_{i}},c_{\bar\beta_{i}}\}$
reflecting an arbitrary sequence of quarks and
anti-quarks in the color-ordered amplitude.  Again in our convention
one negative-helicity gluon is always located on leg $n$\footnote{We
comment in section~\ref{proof} on how to circumvent this problem for
the case without a single negative-helicity gluon.}.

The general N${}^{\text{p}}$MHV tree-amplitude for such a configuration then
reads
\begin{equation}\boxed{
\begin{array}{rl}
A^{\text{N${}^{\text{p}}$MHV}}_{(q\bar q)^{k},n}
(c_{0},\ldots,c_{\alpha_{1}},\ldots ,c_{\bar \beta_{1}}, 
\ldots , c_{\alpha_{k}},
\ldots, c_{\bar \beta_{k}},\ldots, c_{p+k},n)&= 
\displaystyle\frac{\delta^{(4)}(p)}{\ang{1}{2}\ang{2}{3}\ldots\ang{n}{1}}
\times\\
&\hspace{-8cm} \times\quad\text{sign}(\tau)\!\!\!
\displaystyle \sum_{\substack{\text{all paths}\\\text{of length $p$}}}
  \displaystyle \left(\prod_{i=1}^{p}
{\tilde R}^{\{L_{i}\};\{U_{i}\}}_{n;\{I_{i}\}; a_{i}b_{i}}\right)
\, \det \Bigl(\Xi_{n}^{\text{path}}
\Bigr|_{q}\Bigr)^{3}\,
\det \Bigl(\Xi_{n}^{\text{path}}
(\bar{q}\leftrightarrow q)\Bigr|_{\bar{q}}\Bigr)\,.
\end{array}
}
\label{masterquark}
\end{equation}
Here $\text{sign}(\tau)$ is the sign produced by permuting the
quark and anti-quark legs into the alternating order
$\{c_{\alpha_{1}},c_{\bar\beta_{1}},c_{\alpha_{2}},c_{\bar\beta_{2}},
\ldots, c_{\alpha_{k}},c_{\bar\beta_{k}}\}$.

Remarkably, the only difference from the pure gluon amplitudes is a change in
the determinant factors of the path-matrix $\Xi^{\text{path}}_{n}$. 
With $2+p+k$ `special' legs the path-matrix associated to 
\eqn{speciallegs} is now a $(p+1)\times(p+1+k) $ matrix of the form
\begin{equation}
\Xi_{n}^{\text{path}} :=
\left (
\begin{array}{lllllll}
\vev{nc_{0}}  & \ldots &\vev{nc_{\alpha_{i}}} &
\ldots &\vev{nc_{\bar\beta_{i}}} & \ldots & \vev{nc_{p+k}} \cr\cr
(\Xi_{n})^{c_{0}}_{a_{1}b_{1}} & \ldots &
(\Xi_{n})^{c_{\alpha_{i}}}_{a_{1}b_{1}} & 
\ldots &
(\Xi_{n})^{c_{\bar\beta_{i}}}_{a_{1}b_{1}}&
\ldots & (\Xi_{n})^{c_{p}}_{a_{1}b_{1}} \cr\cr
 (\Xi_{n})^{c_{0}}_{\{I_{2}\};a_{2}b_{2}} 
 &\ldots & (\Xi_{n})^{c_{\alpha_{i}}}_{\{I_{2}\};a_{2}b_{2}}  
 &\ldots & (\Xi_{n})^{c_{\bar\beta_{i}}}_{\{I_{2}\};a_{2}b_{2}} &
 \ldots & (\Xi_{n})^{c_{p}}_{\{I_{2}\};a_{2}b_{2}} \cr
 \,\,\,\,\vdots & &\,\,\,\, \vdots & &\,\,\,\,\vdots && \,\,\,\,\vdots \cr
 (\Xi_{n})^{c_{0}}_{\{I_{p}\};a_{p}b_{p}} &
 \ldots & (\Xi_{n})^{c_{\alpha_{i}}}_{\{I_{p}\};a_{p}b_{p}} & 
  \ldots & (\Xi_{n})^{c_{\bar\beta_{i}}}_{\{I_{p}\};a_{p}b_{p}} &
  \ldots & (\Xi_{n})^{c_{p}}_{\{I_{p}\};a_{p}b_{p}} \cr
\end{array}\right )
\label{pm2}
\end{equation}
The notation $\Xi^{\text{path}}_{n}|_{q}$ in \eqn{masterquark} now refers 
to the elimination of all the quark columns (the $c_{\alpha_{i}}$'s)
in the path-matrix and
$\Xi^{\text{path}}_{n}(\bar{q}\leftrightarrow q)|_{\bar{q}}$
denotes the matrix with all the anti-quark columns removed 
(the $c_{\bar\beta_{i}}$'s) after quark and anti-quark columns have been
interchanged, {\it i.e.} $c_{\bar\beta_{i}}\leftrightarrow c_{\alpha_i}$. 
The removal of $k$ columns is
of course necessary in order to turn the $(p+1+k)\times (p+1)$ matrix
$\Xi^{\text{path}}_{n}$ into square form, so that the determinant is
defined.

Let us again spell out some of the lower $p$ amplitudes explicitly to
clarify the formula~\eqref{masterquark}:
\begin{align}
%
%
A^{\text{MHV}}_{q\bar q,n}(c_{\alpha_{1}},c_{\bar\beta_{1}},n) &=
\frac{\delta^{(4)}(p)}{\vev{12}\vev{23}\ldots\vev{n1}}
 \cdot \vev{nc_{\bar\beta_{1}}}^{3}
\cdot \vev{nc_{\alpha_{1}}}\\
%
%
A^{\text{NMHV}}_{q\bar q,n}(c_{0},c_{\alpha_{1}},c_{\bar\beta_{1}},n) &= 
\frac{\delta^{(4)}(p)}{\vev{12}\vev{23}\ldots\vev{n1}} 
\sum_{2\leq a_{1}<b_{1}\leq n-1} 
  {\tilde R}_{n;a_{1}b_{1}}\cdot\nn\\& \qquad
\left |
\begin{array}{llll}
\vev{nc_{0}}  & \vev{nc_{\bar\beta_{1}}} \cr
(\Xi_{n})^{c_{0}}_{a_{1}b_{1}} & (\Xi_{n})^{c_{\bar\beta_{1}}}_{a_{1}b_{1}} \cr
\end{array}\right |^{3}\cdot
\left |
\begin{array}{llll}
\vev{nc_{0}}  & \vev{nc_{\alpha_{1}}} \cr
(\Xi_{n})^{c_{0}}_{a_{1}b_{1}} & (\Xi_{n})^{c_{\alpha_{1}}}_{a_{1}b_{1}}  \cr
\end{array}\right |\\
%
%
A^{\text{NMHV}}_{(q\bar q)^{2},n}
(c_{\alpha_{1}},c_{\bar\beta_{1}},c_{\alpha_{2}},c_{\bar\beta_{2}},n) &= 
\frac{\delta^{(4)}(p)}{\vev{12}\vev{23}\ldots\vev{n1}} 
\sum_{2\leq a_{1}<b_{1}\leq n-1} 
  {\tilde R}_{n;a_{1}b_{1}}\cdot \nn\\& \qquad
\left |
\begin{array}{llll}
\vev{nc_{\bar\beta_{1}}}  & \vev{nc_{\bar\beta_{2}}} \cr
(\Xi_{n})^{c_{\bar\beta_{1}}}_{a_{1}b_{1}} 
& (\Xi_{n})^{c_{\bar\beta_{2}}}_{a_{1}b_{1}}  \cr
\end{array}\right |^{3}\cdot
\left |
\begin{array}{llll}
\vev{nc_{\alpha_{1}}}  & \vev{nc_{\alpha_{2}}} \cr
(\Xi_{n})^{c_{\alpha_{1}}}_{a_{1}b_{1}}
& (\Xi_{n})^{c_{\alpha_{2}}}_{a_{1}b_{1}}  \cr
\end{array}\right |\\
%
%
A_{q\bar q,n}^{\text{N${}^2$MHV}}
(c_{\alpha_{1}},c_{1},c_{\bar\beta_{1}},c_{2},n)&=
\frac{\delta^{(4)}(p)}{\vev{12}\vev{23}\ldots\vev{n1}}\sum_{2\leq a_1<b_1<n}
{\tilde R}_{n;a_1b_1}\cdot \nn \\ \Biggl[\sum_{a_1+1\leq a_2<b_2\leq b_1}
{\tilde R}^{0;a_1 b_1}_{n;b_{1}a_{1};a_{2}b_{2}} 
\cdot & \left |
\begin{array}{llll}
\vev{nc_{1}}  & \vev{nc_{\bar\beta_1}} &  \vev{nc_{2}} \cr
(\Xi_{n})^{c_{1}}_{a_{1}b_{1}} & (\Xi_{n})^{c_{\bar\beta_1}}_{a_{1}b_{1}} &
(\Xi_{n})^{c_{2}}_{a_{1}b_{1}} \cr
(\Xi_{n})^{c_{1}}_{b_{1},a_{1};a_{2}b_{2}}
& (\Xi_{n})^{c_{\bar\beta_1}}_{b_{1},a_{1};a_{2}b_{2}} &
(\Xi_{n})^{c_{2}}_{b_{1},a_{1};a_{2}b_{2}} \cr
\end{array}\right |^{3}\cdot\nn\\&
\left |
\begin{array}{llll}
\vev{nc_{1}}  & \vev{nc_{\alpha_1}} &  \vev{nc_{2}} \cr
(\Xi_{n})^{c_1}_{a_{1}b_{1}} & (\Xi_{n})^{c_{\alpha_1}}_{a_{1}b_{1}} &
(\Xi_{n})^{c_{2}}_{a_{1}b_{1}} \cr
(\Xi_{n})^{c_1}_{b_{1},a_{1};a_{2}b_{2}}
& (\Xi_{n})^{c_{\alpha_1}}_{b_{1},a_{1};a_{2}b_{2}} &
(\Xi_{n})^{c_2}_{b_{1},a_{1};a_{2}b_{2}} \cr
\end{array}\right | \nn\\
\, 
+\sum_{b_{1}\leq a_{2},b_{2}<n} {\tilde R}^{a_{1}b_{1};0}_{n;a_{2}b_{2}}
\cdot &\left |
\begin{array}{llll}
\vev{nc_{1}}  & \vev{nc_{\bar\beta_1}} &  \vev{nc_{2}} \cr
(\Xi_{n})^{c_{1}}_{a_{1}b_{1}} & (\Xi_{n})^{c_{\bar\beta_1}}_{a_{1}b_{1}} &
(\Xi_{n})^{c_{2}}_{a_{1}b_{1}} \cr
(\Xi_{n})^{c_{1}}_{a_{2}b_{2}} & (\Xi_{n})^{c_{\bar\beta_1}}_{a_{2}b_{2}} &
(\Xi_{n})^{c_{2}}_{a_{2}b_{2}} \cr
\end{array}\right |^{3}\, 
\left |
\begin{array}{llll}
\vev{nc_{1}}  & \vev{nc_{\alpha_1}} &  \vev{nc_{2}} \cr
(\Xi_{n})^{c_{1}}_{a_{1}b_{1}} & (\Xi_{n})^{c_{\alpha_1}}_{a_{1}b_{1}} &
(\Xi_{n})^{c_{2}}_{a_{1}b_{1}} \cr
(\Xi_{n})^{c_{1}}_{a_{2}b_{2}} & (\Xi_{n})^{c_{\alpha_1}}_{a_{2}b_{2}} &
(\Xi_{n})^{c_{2}}_{a_{2}b_{2}} \cr
\end{array}\right |\,\Biggr] \\
%
%
A_{(q\bar q)^{2},n}^{\text{N${}^2$MHV}}
(c_{\alpha_{1}},c_{1},c_{\bar\beta_{1}},c_{\alpha_2},
c_{\bar\beta_2},n)&=
\frac{\delta^{(4)}(p)}{\vev{12}\vev{23}\ldots\vev{n1}}\sum_{2\leq a_1<b_1<n}
{\tilde R}_{n;a_1 b_1}\cdot \nn \\ \Biggl[\sum_{a_1+1\leq a_2<b_2\leq b_1}
{\tilde R}^{0;a_1 b_1}_{n;b_{1}a_{1};a_{2}b_{2}} 
\cdot & \left |
\begin{array}{llll}
\vev{nc_{1}}  & \vev{nc_{\bar\beta_1}} &  \vev{nc_{\bar\beta_2}} \cr
(\Xi_{n})^{c_{1}}_{a_{1}b_{1}} & (\Xi_{n})^{c_{\bar\beta_1}}_{a_{1}b_{1}} &
(\Xi_{n})^{c_{\bar\beta_2}}_{a_{1}b_{1}} \cr
(\Xi_{n})^{c_{1}}_{b_{1},a_{1};a_{2}b_{2}}
& (\Xi_{n})^{c_{\bar\beta_1}}_{b_{1},a_{1};a_{2}b_{2}} &
(\Xi_{n})^{c_{\bar\beta_2}}_{b_{1},a_{1};a_{2}b_{2}} \cr
\end{array}\right |^{3}\cdot\nn\\&
\left |
\begin{array}{llll}
\vev{nc_{1}}  & \vev{nc_{\alpha_1}} &  \vev{nc_{\alpha_2}} \cr
(\Xi_{n})^{c_{1}}_{a_{1}b_{1}} & (\Xi_{n})^{c_{\alpha_1}}_{a_{1}b_{1}} &
(\Xi_{n})^{c_{\alpha_2}}_{a_{1}b_{1}} \cr
(\Xi_{n})^{c_{1}}_{b_{1},a_{1};a_{2}b_{2}}
& (\Xi_{n})^{c_{\alpha_1}}_{b_{1},a_{1};a_{2}b_{2}} &
(\Xi_{n})^{c_{\alpha_2}}_{b_{1},a_{1};a_{2}b_{2}} \cr
\end{array}\right | \nn\\
\, 
+\sum_{b_{1}\leq a_{2}<b_{2}<n} {\tilde R}^{a_{1}b_{1};0}_{n;a_{2}b_{2}}
\cdot &\left |
\begin{array}{llll}
\vev{nc_{1}}  & \vev{nc_{\bar\beta_1}} &  \vev{nc_{\bar\beta_2}} \cr
(\Xi_{n})^{c_{1}}_{a_{1}b_{1}} & (\Xi_{n})^{c_{\bar\beta_1}}_{a_{1}b_{1}} &
(\Xi_{n})^{c_{\bar\beta_2}}_{a_{1}b_{1}} \cr
(\Xi_{n})^{c_{1}}_{a_{2}b_{2}} & (\Xi_{n})^{c_{\bar\beta_1}}_{a_{2}b_{2}} &
(\Xi_{n})^{c_{\bar\beta_2}}_{a_{2}b_{2}} \cr
\end{array}\right |^{3}\, 
\left |
\begin{array}{llll}
\vev{nc_{1}}  & \vev{nc_{\alpha_1}} &  \vev{nc_{\alpha_2}} \cr
(\Xi_{n})^{c_{1}}_{a_{1}b_{1}} & (\Xi_{n})^{c_{\alpha_1}}_{a_{1}b_{1}} &
(\Xi_{n})^{c_{\alpha_2}}_{a_{1}b_{1}} \cr
(\Xi_{n})^{c_{1}}_{a_{2}b_{2}} & (\Xi_{n})^{c_{\alpha_1}}_{a_{2}b_{2}} &
(\Xi_{n})^{c_{\alpha_2}}_{a_{2}b_{2}} \cr
\end{array}\right |\,\Biggr] 
\end{align}
This completes our examples. Some explicit formulae with the
determinants expanded out may be found in appendix \ref{explicit2}.
The formula (\ref{masterquark}) is implemented in {\tt GGT} by the function
{\tt GGTfermionS}.  See appendix \ref{GGT} for the documentation.

Note that the master formula~\eqref{masterglue} reduces as it should
to the pure-gluon scattering result~\eqref{masterglue} for a zero number of
quark-anti-quark pairs, $k\to 0$.  In that case no column removals are to
be performed and the determinants combine to the power four.


\section{All gluon-gluino tree amplitudes in
\texorpdfstring{${\cal N}=4$}{N=4} super Yang-Mills}
\label{sect:all-flavor}

The color-ordered tree amplitudes with fermions presented above were
special in the sense that they apply both to massless QCD as well as
${\cal N}=4$ super Yang-Mills, due to the single-flavor choice which
suppresses intermediate scalar exchange as argued in section
\ref{sect-SYMtoQCD}. We now state the master formula for general
gluino and gluon tree amplitudes in the ${\cal N}=4$ super Yang-Mills
theory from which the above expressions arise. Through specific
choices of external flavor configurations, however, this master
formula may be used to produce color-ordered gluon and quark trees in
massless QCD, as was discussed in section \ref{sect-SYMtoQCD}.

Similar to the notation in section \ref{sect:mono-flavor}, for a
general $g^{n-2k}(q\bar q)^{k}$ amplitude with arbitrary flavor
assignments to the `quarks'\footnote{We refer to the gluinos in this
and the following sections as `quarks' in order to not introduce 
new terminology beyond that introduced already in sections 4 and 5.} 
we have $2+p+k$ `special' legs (negative-helicity gluon, quark or
anti-quark), whose position amongst the $n$ legs we parametrize by the set
\begin{equation}
(c_{0},\ldots,c^{A_{1}}_{\alpha_{1}},\ldots ,
c^{B_{1}}_{\bar \beta_{1}} \ldots , c^{A_{k}}_{\alpha_{k}},
\ldots, c^{B_{k}}_{\bar \beta_{k}},\ldots, c_{p+k},n)\, .
\label{speciallegsSYM}
\end{equation}
Again the configuration of quarks and anti-quarks inside the gluon
background may be arbitrary while the sets $\{\alpha_{i}\}$ and
$\{\bar\beta_{i}\}$ are ordered. The general $g^{n-2k}(q\bar q)^{k}$
amplitude with $(2+p-k)$ negative-helicity gluons is then
expressed in terms of the $\tilde R$-functions and the path-matrix
$\Xi^{\text{path}}_{(q\bar q)^{k},n}$ defined above. It reads
\begin{equation}\!\!\!\!\!\!\boxed{
\begin{array}{rl}
 A^{\text{N${}^{\text{p}}$MHV}}_{(q\bar q)^{k},n}
(c_{0},\ldots,c_{\alpha_{i}}^{A_i},\ldots  ,
c_{\bar \beta_{j}}^{B_j} \ldots , & c_{p+k},n)=
\displaystyle
\frac{\delta^{(4)}(p)\sign(\tau)}{\ang{1}{2}\ang{2}{3}\dots \ang{n}{1}}\times \\
 \times &\displaystyle\sum_{\substack{\text{all paths}\\\text{of length } p}} 
\,\left(\prod_{i=1}^{p}
\tilde{R}^{\{L_i\};\{U_i\}}_{n;\{I_i\};a_i b_i}\right) \,
\left(\det \Xi^{\text{path}}_{n}\Bigr
\rvert_{q}\right)^{4-k} \times \\ \times
& \displaystyle
\sum_{\sigma \in S_k}
\sign(\sigma)\prod_{i=1}^{k}\delta^{A_iB_{\sigma(i)}}
\det\left(\Xi^{\text{path}}_{n}
\Bigr\rvert_{q}(\bar{\beta}_{\sigma(i)}
\leftrightarrow\alpha_i)\right)\, .
\end{array}
}
\label{master_quark_SYM_original}
\end{equation}
Here the notation
$\Xi^{\text{path}}_{n}\rvert_{q}$ 
refers to the path-matrix~(\ref{pm2}) with all `quark' columns
$\{c_{\alpha_{i}}\}$ removed, whereas 
$\Xi^{\text{path}}_{n}
\rvert_{q}(\bar{\beta}_{i}
\rightarrow\alpha_i)$ denotes the same path-matrix where
the `anti-quark' column $c_{\bar\beta_{j}}$ is replaced by one of the
previously-removed `quark' columns $c_{\alpha_{i}}$.  Also $\sign(\tau)$
is the sign of the permutation for bringing the initial color-ordering
of the fermionic legs into the canonical order
$\{c_{\alpha_{1}},c_{\bar\beta_{1}},c_{\alpha_{2}},
c_{\bar\beta_{2}},\ldots, c_{\alpha_{k}},c_{\bar\beta_{k}}\}$
and $\sign(\sigma)$ is the sign of the permutation of $\sigma$.
This formula is implemented in {\tt GGT} by the function {\tt GGTfermion}.

\section{Proof of the master formula}
\label{proof}

In this section we prove the master formula~(\ref{master_quark_SYM_original})
for a general ${\cal N}=4$ super Yang-Mills tree amplitude with external
gluons and gluinos of arbitrary flavor, as well as the more compact
single-flavor expression~(\ref{masterquark}) and the 
pure gluon expression~(\ref{masterglue}) as sub-cases.

Amplitudes in ${\cal N}=4$ super Yang-Mills are very efficiently
expressed in terms of a superwave function $\Phi$ which collects all on-shell
states of the PCT self-conjugate massless ${\cal N}=4$ multiplet, with
the help of the Grassmann variables $\eta^{A}$ with $A=1,2,3,4$ of the
$su(4)$ R-symmetry,
\begin{align} \label{super-wave}
  \Phi(\lambda,\tilde{\lambda},\eta)
 =& g^{+}(\lambda,\tilde{\lambda})
 + \eta^A \,{\tilde g}_A(\lambda,\tilde{\lambda})
 + \frac{1}{2}\eta^A \eta^B\, \phi_{AB}(\lambda,\tilde{\lambda})
  + \frac{1}{3!}\eta^A\eta^B\eta^C \epsilon_{ABCD} \,
  {\bar{\tilde g}}^{D}(\lambda,\tilde{\lambda}) \nonumber \\
  &\  + \frac{1}{4!}\eta^A\eta^B\eta^C \eta^D \epsilon_{ABCD} \,
  g^{-}(\lambda,\tilde{\lambda})\,.
\end{align}
Here $g^{\pm}$ are the $\pm 1$ helicity gluons, ${\tilde g}_{A}$ and
$\bar{\tilde g}^{A}$ the four flavor $\pm \frac12$ helicity gluinos,
and $\phi_{AB}$ the six real $0$ helicity scalar states.
The Grassmann variables
$\eta$ carry helicity $+\tfrac{1}{2}$ so that the whole multiplet
carries helicity $+1$.

We can write the amplitudes in the on-shell superspace with
coordinates $(\lambda_i,\tilde\lambda_i,\eta_i)$~\cite{%
Nair:1988bq,Witten:2003nn,Georgiou:2004by},
\be\label{super-amplitude}
{\cal A}_n\big(\lambda_i,\tilde{\lambda}_i,\eta_i \big)
= \mathcal{A}\left( \Phi_1 \ldots \Phi_n \right)\,.
\ee
Since the helicity of each supermultiplet $\Phi_i$ is $1$ the amplitude obeys
\be
\label{helicity}
h_i \mathcal{A}_n(\lambda_i,\tilde\lambda_i,\eta_i)
= \mathcal{A}_n(\lambda_i,\tilde\lambda_i,\eta_i)\,,
\ee
where $h_{i}$ is the helicity operator for the $i$th leg. The component
field amplitudes are then obtained by projecting upon the relevant terms
in the $\eta_{i}$ expansion of the super-amplitude via
\be
g^{+}_{i} \to \eta^{A}_{i}=0\, , \quad
g^{-}_{i} \to \int d^{4}\eta_{i}
= \int d\eta_{i}^{1}\,d\eta_{i}^{2}\, d\eta_{i}^{3}\,d\eta_{i}^{4}\,, \quad
\tilde g_{i,A} \to \int d\eta^{A}\, ,\quad  \bar{\tilde g}^{A}_{i} 
\to -\int d^{4}\eta_{i} \, \eta_{i}^{A}\, .
\ee
Note that the super-amplitude ${\cal A}_n$ has a cyclic symmetry
that can lead to many different but equivalent expressions in practice.
In order to obtain compact expressions for component amplitudes 
often a judicious choice of this cyclic freedom can be
made~\cite{Drummond:2008cr}.

The general solution for tree super-amplitudes
of Drummond and one of the present authors~\cite{Drummond:2008cr}
takes the compact form
\begin{align}
{\cal A}^{\text{N${}^{p}$MHV}}_{n}&=
\frac{\delta^{(4)}(p)\, \delta^{(8)}(q)}{\vev{12}\vev{23}\ldots\vev{n1}} 
 \sum_{\text{all paths of length $p$}}
  \displaystyle 1\cdot{ R}_{n,a_{1}b_{1}}\cdot
{R}^{\{L_{2}\};\{U_{2}\}}_{n,\{I_{2}\}, a_{2}b_{2}}\cdot \ldots \cdot 
{R}^{\{L_{p}\};\{U_{p}\}}_{n,\{I_{p}\}, a_{p}b_{p}} \, ,
\label{mastersuper}
\end{align}
where $q^{\alpha\, A}=\sum_{i=1}^{n}\lambda^{\alpha}_{i}\, \eta^{A}_{i}$
is the total conserved fermionic momentum, and the dual superconformal
$R$-invariant is
\begin{equation}
R_{n;\{I\};ab} = {\tilde R}_{n;\{I\};ab}\,
\delta^{(4)}\left(\sum_{i=1}^{n}\Xi_{n;\{I\};ab}(i)\,\eta_{i}\right) \,,
\label{Rdef}
\end{equation}
in the notation of the previous sections. We now wish to project
this result in on-shell superspace onto the relevant components for a
general $g^{n-2k}(q\bar q)^{k}$ amplitude. For this purpose we set all
of the $\eta_{i}$ associated to positive-helicity gluon states to
zero.  This leaves us with the $p+2+k$ remaining Grassmann numbers
$\eta_{c_{i}}$ associated to the `special' legs of helicities $-1$ and
$\pm\frac{1}{2}$. To project onto a negative-gluon state at position
$i$ one simply has to integrate \eqn{mastersuper} against $\int
d^{4}\eta_{i}$. Similarly, to project to a quark or anti-quark state
at position $i$ of flavor $A_{i}$ one integrates \eqn{mastersuper}
against $\int d\eta_{i}^{A_{i}}$ or $-\!\int d^{4}\eta_{i}\,\eta_{i}^{A_{i}}$.
All integrations have to be in color order.

In accord with our convention above the leg $n$ is chosen to be a
negative-helicity gluon state, or an anti-quark if there are no
negative-helicity gluons.  This is a convenient choice because the only
dependence of the super-amplitude on $\eta_n$ is through the total 
fermionic momentum conserving delta function, which can be written as
\begin{equation}\label{delta8}
\delta^{{(8)}}(q)=\delta^{(8)}\left(\sum_{i=0}^{p+k}
\lambda_{c_i}\, \eta_{c_{i}}+\lambda_{n}\,
\eta_{n}\right) = \delta^{(4)}\left(\sum_{i=1}^{p+k}
\frac{\ang{c_0}{c_i}}{\ang{c_{0}}{n}}\, \eta_{c_{i}}
+\eta_{n}\right)\,
\delta^{(4)}\left(\sum_{i=0}^{p+k}\ang{n}{c_{i}}\, \eta_{c_{i}}\right)\,.
\end{equation}
For each path in \eqn{mastersuper} the four-dimensional Grassmann 
delta functions in \eqn{delta8}, together with the $p$ delta functions
arising from the $R$-invariants~\eqref{Rdef}, may be written as
\begin{multline}
\prod_{i=0}^{p+1}\delta^{(4)}
\left( \sum_{j=0}^{p+k}\left(\Xi^{\text{path}}_{n}\right)_{ij}
\eta_{c_j}\right) := \\
\delta^{(4)}\left(\sum_{i=1}^{p+k}\frac{\ang{c_0}{c_i}}{\ang{c_{0}}{n}}\,
\eta_{c_{i}}+\eta_{n}\right)\,
\delta^{(4)}\left(\sum_{i=0}^{p+k}\ang{n}{c_{i}}\, \eta_{c_{i}}\right)
\prod_{i=1}^{p}\, \delta^{(4)}\left(\sum_{j=0}^{p+k}\Xi_{n;\{I_i\};a_ib_i}(c_j)
\, \eta_{c_{j}}\right)\,,
\label{eq:DEFpathMatrix}
\end{multline}
with the $(p+2)\times(p+k+2)$ path-matrix
$\Xi^{\text{path}}_{n}$.  If we have a negative-helicity
gluon at position $n$, the $\eta_n$ integration is trivial and we can
drop the trial $\eta_n$ column and the row determined by the first
delta function in \eqn{eq:DEFpathMatrix}, ending up with the
$(p+1)\times(p+k+1)$ path-matrix given in eq.~\eqref{pm2}. For the
sake of readability we will drop the labels on the path-matrix in what
follows, just denoting it by $\Xi$, and assume a negative-helicity gluon
at position $n$. The projection to the general $g^{n-2k}(q\bar q)^{k}$
amplitude of \eqn{masterglue}, with quarks of flavor $A_i$ at
positions $c_{\alpha_i}$ and anti-quarks of flavor $B_j$ at positions
$c_{\bar{\beta}_j}$,
\begin{equation*}
A^{\text{N${}^{\text{p}}$MHV}}_{(q\bar q)^{k},n}
(c_{0},\ldots,c_{\alpha_{i}}^{A_i},\ldots ,c_{\bar \beta_{j}}^{B_j}, 
\ldots , c_{p+k},n) \, ,
\end{equation*}
is then performed via the Grassmann integrals
\begin{equation}\label{grassmann-integral}
(-1)^k\sign(\tau)\left(\prod_{\substack{j=0\\j
\notin \{\alpha_1,\,\dots\,,\alpha_k\}}}^{p+k}
\int d^{4}\eta_{c_{j}}\right)\,
\left(\prod_{l=1}^{k}
\int d\eta^{A_l}_{c_{\alpha_{l}}}\, \eta^{B_l}_{c_{\bar{\beta}_{l}}}\right)\, 
\prod_{i=0}^{p+1}
\delta^{(4)}\left( \sum_{j=0}^{p+k}\Xi_{i\,j}\eta_{c_j}\right) \,.
\end{equation}
Here $\sign(\tau)$ compensates the minus signs we encountered by
permuting the quark and anti-quark Grassmann variables 
from color order to the canonical order
$\prod_{l=1}^{k}\int d\eta^{A_l}_{c_{\alpha_{l}}}\,
\eta^{B_l}_{c_{\bar{\beta}_{l}}}$.

Let us first consider the pure gluon case, {\it i.e.}~$k=0$. Performing
the change of variables $\eta_{c_i}\rightarrow \Xi^{-1}_{ij}\eta_{c_j}$
immediately gives 
\begin{equation}
 \left(\prod_{j=0}^{p}\int d^{4}\eta_{c_{j}}\right)\,
\prod_{i=0}^{p+1}\delta^{(4)}
\left( \sum_{j=0}^{p}\Xi_{i\,j}\eta_{c_j}\right)=\left(\det \Xi\right)^4\,,
\end{equation}
thereby proving \eqn{masterglue}.  To evaluate the general
integral~\eqref{grassmann-integral} we first perform the $(p+1)$
four-dimensional integrations with respect to the $\eta$'s of the
anti-quarks and gluons by making a change of variables similar to
the pure-gluon case, leading to
\begin{equation}
  \sign(\tau)\left(\det \Xi\bigr\rvert_q \right)^4
\prod_{l=1}^{k}\int d\eta^{A_l}_{c_{\alpha_{l}}}\,
\sum_{\substack{i=0\\i\notin\{\alpha_1,\,\dots\,,\alpha_k\}}}^{p+k}
\left(\Xi\bigr\rvert_q^{-1}\right)_{\bar{\beta_l}\,i}
\sum_{j=1}^k\Xi_{i\,\alpha_j}\eta_{c_{\alpha_j}}^{B_l} \,.
  \label{eq:proofStep1}
\end{equation}
Here $\Xi\bigr\rvert_q$ refers to the elimination of all quark columns
in the path-matrix.  We can simplify the sum over $i$ by making use of some
basic facts of linear algebra.  Namely, given a square matrix
$M=\left(m_{ij}\right)$ with minors $M_{ij}$, its determinant and inverse
can be written as 
\begin{align}
\det M &= \sum_{i} (-1)^{i+j} m_{ij} \det M_{ij}&\text{and}&&
\left(M^{-1}\right)_{ij}&=(-1)^{i+j}\frac{\det M_{ji}}{\det M}\,.
\end{align}
Hence, \eqn{eq:proofStep1} simplifies to
\begin{equation}
\sign(\tau)\left(\det \Xi\bigr\rvert_q \right)^{4-k}\,
\prod_{l=1}^{k}\int d\eta^{A_l}_{c_{\alpha_{l}}}\,
\sum_{j=1}^k
\det\left(\Xi\bigr\rvert_q(\bar{\beta}_l\rightarrow\alpha_j)\right)
\eta_{c_{\alpha_j}}^{B_l}\,,
\end{equation}
where $\Xi\bigr\rvert_q(\bar{\beta}_l\rightarrow\alpha_j)$ 
denotes the replacement of an anti-quark column by a quark column.
The remaining integrations are straightforward and give
\begin{equation}
 \sign(\tau)\left(\det \Xi\bigr\rvert_q \right)^{4-k}
\,\sum_{\sigma \in S_k}\sign(\sigma)
\prod_{i=1}^{k}\delta^{A_iB_{\sigma(i)}}
\det\left(\Xi\bigr\rvert_q
(\bar{\beta}_{\sigma(i)}\rightarrow\alpha_i)\right)\,.
\end{equation}
The general $\mathcal{N}=4$ super Yang-Mills $g^{n-2k}(q\bar q)^{k}$
amplitude is therefore
\begin{multline}\label{master_quark_SYM}
 A^{\text{N${}^{\text{p}}$MHV}}_{(q\bar q)^{k},n}
(c_{0},\ldots,c_{\alpha_{i}}^{A_i},\ldots ,c_{\bar \beta_{j}}^{B_j}
 \ldots , c_{p+k},n)
=\frac{\delta^{(4)}(p)\sign(\tau)}
{\ang{1}{2}\ang{2}{3}\dots \ang{n}{1}}\times\\
\sum_{\substack{\text{all paths}\\\text{of length } p}}
\left(\prod_{i=1}^p\tilde{R}^{L_i;R_i}_{n;\{I_i\};a_i b_i}\right)
\left(\det \Xi\bigr\rvert_q \right)^{4-k}
\sum_{\sigma \in S_k}\sign(\sigma)
\prod_{i=1}^{k}\delta^{A_iB_{\sigma(i)}}
\det\left(\Xi\bigr\rvert_q(\bar{\beta}_{\sigma(i)}\rightarrow\alpha_i)\right)
\,.
\end{multline}
Note that during the derivation of this formula we assumed that there is
at least one negative-helicity gluon.  The only change in the case $k=p+2$
is that the $\eta_n$ integration is no longer trivial and we put a gluino
at position $n$. Hence, the path
matrix has the size $(p+2)\times(2p+4)$ and \eqn{master_quark_SYM}
still holds as its derivation did not depend on the matrix dimensions.  
Using \eqref{eq:DEFpathMatrix} the path matrix \eqref{Pathmatrix1} then generalizes to the form
\begin{equation}
\label{Pathmatrix2}
\Xi
:=
\left (
\begin{array}{llllll}
0 & \frac{\vev{c_{0}c_{1}}}{\vev{c_{0}n}} & 
\frac{\vev{c_{0}c_{2}}}{\vev{c_{0}n}} & \ldots & 
\frac{\vev{c_{0}c_{p+k}}}{\vev{c_{0}n}} & 1 \cr \cr
\vev{nc_{0}}  & \vev{nc_{1}} 
& \vev{nc_{2}} &\ldots & \vev{nc_{p+k}} & 0 \cr\cr
(\Xi_{n})^{c_{0}}_{a_{1}b_{1}} & (\Xi_{n})^{c_{1}}_{a_{1}b_{1}} 
& (\Xi_{n})^{c_{2}}_{a_{1}b_{1}} 
                             & \ldots & (\Xi_{n})^{c_{p+k}}_{a_{1}b_{1}} & 0\cr\cr
 (\Xi_{n})^{c_{0}}_{\{I_{2}\};a_{2}b_{2}}
    & (\Xi_{n})^{c_{1}}_{\{I_{2}\};a_{2}b_{2}}
    & (\Xi_{n})^{c_{2}}_{\{I_{2}\};a_{2}b_{2}}
                        & \ldots & (\Xi_{n})^{c_{p+k}}_{\{I_{2}\};a_{2}b_{2}} & 0\cr
\,\,\,\, \vdots & \,\,\,\,\vdots & & & \,\,\,\,\vdots & \vdots \cr
 (\Xi_{n})^{c_{0}}_{\{I_{p}\};a_{p}b_{p}}
    & (\Xi_{n})^{c_{1}}_{\{I_{p}\};a_{p}b_{p}}
      & (\Xi_{n})^{c_{2}}_{\{I_{p}\};a_{p}b_{p}}
                      & \ldots & (\Xi_{n})^{c_{p+k}}_{\{I_{p}\};a_{p}b_{p}}&0 \cr
\end{array}\right ) \ .
\end{equation}
Generally the amplitudes take a more compact form if the gluino at position
$n$ is taken to be of helicity $-1/2$.
Several explicit formulas for the MHV and NMHV cases can
be found in Appendix \ref{explicit2}. In particular
Appendix \ref{NMHVsixfermions}
discusses a case without a negative helicity gluon at position $n$.

As we are interested in QCD tree amplitudes, we need to decouple
possible intermediate scalar states arising from the Yukawa couplings
$\tilde g_{A}\tilde g_{B} \phi^{AB}$ in the ${\cal N}=4$ super Yang-Mills
Lagrangian.   As discussed in section~\ref{sect-SYMtoQCD},
one case in which this can be achieved (although not the only one needed
for QCD) is when the external fermion states all have the same flavor,
due to the anti-symmetry of $\phi^{AB}=\epsilon^{ABCD}\,\phi_{CD}$.
For this case, we specialize to $A_{i}=B_{i}=A$ for all external fermion
legs $i$ in our master formula~\eqref{master_quark_SYM}, and perform
the sum over permutations explicitly, yielding
\begin{multline}
 A^{\text{N${}^{\text{p}}$MHV}}_{(q\bar q)^{k},n}
(c_{0},\ldots,c_{\alpha_{i}},\ldots ,c_{\bar \beta_{j}}\ldots , c_{p+k},n)=\\
\frac{\delta^{(4)}(p)\sign(\tau)}{\ang{1}{2}\ang{2}{3}\dots \ang{n}{1}}
\sum_{\substack{\text{all paths}\\\text{of length } p}}
\left(\prod_{i=1}^p\tilde{R}^{L_i;R_i}_{n;\{I_i\};a_i b_i}\right)
\det \Xi(q\leftrightarrow\bar{q})\bigr\rvert_{\bar{q}}
\left(\det \Xi\bigr\rvert_q \right)^{3}\,,
\end{multline}
which reproduces~\eqn{masterquark}.

\subsection*{Acknowledgments}

We thank Zvi Bern, Benedikt Biedermann, Jake Bourjaily, Harald Ita,
Kemal Ozeren
and Peter Uwer for helpful discussions.  Several figures in this paper were
made with {\sc Jaxodraw}~\cite{Binosi:2003yf,Binosi:2008ig}, based on
{\sc Axodraw}~\cite{Vermaseren:1994je}. 
This work was supported by the Volkswagen Foundation, and by the US
Department of Energy under contract DE-AC02-76SF00515.

\appendix

\section{Explicit formulae for gluon trees}\label{explicit}
Here we explicitly apply our formula~\eqref{masterglue} to the NMHV and
NNMHV cases.
\subsection{NMHV amplitudes}
Without loss of generality, we take the negative-helicity gluons to be
at positions $c_{0}, c_{1}, n$ with $c_{0}<c_{1}$. In the NMHV case
only one path in figure~\ref{rootedtree} contributes and the
path-matrix is a $2\times 2$ matrix whose determinant we denote by
$D_{n,a_1b_1}^{c_0c_1}$. Hence, the NMHV gluon amplitude is given by
\begin{equation}
A^{\text{NMHV}}_{n}(c_{0}^-,c_{1}^-,n^-) = 
\frac{\delta^{(4)}(p)}{\ang{1}{2}\ldots\ang{n}{1}} 
\sum_{2\leq a_{1}<b_{1}\leq n-1} 
  {\tilde R}_{n;a_{1}b_{1}}\cdot
\left (
D_{n,a_1b_1}^{c_0c_1}\right )^{4}
\end{equation}
where the determinant of the path-matrix is given explicitly by
\begin{equation}
D_{n,st}^{ab}:=\left |
\begin{array}{llll}
\ang{n}{a}  & \ang{n}{b} \cr
(\Xi_{n})^{a}_{st} & (\Xi_{n})^{b}_{st}  \cr
\end{array}\right |\stackrel{a<b}{=}
\begin{cases}
\ang{n}{a}\langle n t s |b\rangle & \qquad a<s\leq b<t \,, \\
\ang{n}{a}\ang{b}{n}x_{st}^2 & \qquad a<s<t\leq b \,, \\
\ang{b}{a}\langle n t s | n\rangle & \qquad s\leq a,b<t \,, \\
\ang{n}{b}\langle n s t | a\rangle & \qquad s\leq a<t\leq b \,.
\end{cases}
\label{eq:Dsymbol}
\end{equation}
For $a>b$ one can use the antisymmetry of the determinant,
$D_{n,st}^{ab} = -D_{n,st}^{ba}$.  \Eqn{eq:Dsymbol} is exactly the
result we already stated in \eqn{intro-nmhv-example}.
This formula is implemented in {\tt GGT} by {\tt GGTnmhvgluon}.
\subsection{\texorpdfstring{N${}^2$MHV}{NNMHV} amplitudes}
The negative-helicity gluons are taken to be $a^-$, $b^-$, $c^-$, $n^-$
with $a<b<c$, without loss of generality. According to
figure~\ref{rootedtree} there are two contributing paths. Denoting the
determinants of their corresponding path-matrices by
$D^{abc}_{\!\scriptscriptstyle1}$ and $D^{abc}_{\!\scriptscriptstyle2}$,
the NNMHV gluon amplitude is given by
\begin{equation}
A_n^{\text{N${}^2$MHV}}(a^-,b^-,c^-,n^-)=
\frac{\delta^{(4)}(p)}{\ang{1}{2}\ldots\ang{n}{1}}\sum_{2\leq a_1<b_1<n}
{\tilde R}_{n;a_1 b_1}\begin{aligned}[t]\cdot\Biggl[
&\sum_{a_1+1\leq a_2<b_2\leq b_1}
{\tilde R}^{0;a_1 b_1}_{n;b_{1}a_{1};a_{2}b_{2}} 
\cdot \left(D^{abc}_{\!\scriptscriptstyle1}\right)^{4}\\
&+\sum_{b_{1}\leq a_{2}<b_{2}<n} {\tilde R}^{a_{1}b_{1};0}_{n;a_{2}b_{2}}
\cdot\left(D^{abc}_{\!\scriptscriptstyle2}\right)^{4}\, \Biggr]\,.
\end{aligned}
\end{equation}
The explicit forms of the determinants of the path-matrices
\begin{equation}
D^{abc}_{\!\scriptscriptstyle 1}(n,a_1,b_1,a_2,b_2):=
\left |
\begin{array}{llll}
\ang{n}{a}  & \ang{n}{b} &  \ang{n}{c} \cr
(\Xi_{n})^{a}_{a_{1}b_{1}} & (\Xi_{n})^{b}_{a_{1}b_{1}} &
(\Xi_{n})^{c}_{a_{1}b_{1}} \cr
(\Xi_{n})^{a}_{b_{1},a_{1};a_{2}b_{2}} 
& (\Xi_{n})^{b}_{b_{1},a_{1};a_{2}b_{2}}
& (\Xi_{n})^{c}_{b_{1},a_{1};a_{2}b_{2}}\cr 
\end{array}\right |
\end{equation}
and
\begin{equation}
D^{abc}_{\!\scriptscriptstyle 2}(n,a_1,b_1,a_2,b_2):=
\left |
\begin{array}{llll}
\ang{n}{a}  & \ang{n}{b} &  \ang{n}{c} \cr
(\Xi_{n})^{a}_{a_{1}b_{1}} & (\Xi_{n})^{b}_{a_{1}b_{1}} &
(\Xi_{n})^{c}_{a_{1}b_{1}} \cr
(\Xi_{n})^{a}_{a_{2}b_{2}} & (\Xi_{n})^{b}_{a_{2}b_{2}} &
(\Xi_{n})^{c}_{a_{2}b_{2}} \cr
\end{array}\right |
\end{equation}
are given by
\begin{equation}
D_1^{abc}
=\begin{cases}
\ang{a}{n}\bracket{nb_1 a_1}{b}\bracket{nb_1 a_1 b_2 a_2}{c}&\qquad a<a_1\leq b,c<b_1\qquad b<a_2\leq c< b_2\\
\ang{n}{a}\bracket{nb_1 a_1}{b}\bracket{nb_1 a_1}{c}x_{a_2b_2}^2&\qquad a<a_1\leq b,c<b_1\qquad b<a_2,b_2\leq c\\
\ang{a}{n}\ang{b}{c}\bracket{nb_1 a_1 a_2 b_2}{nb_1a_1}&\qquad a<a_1\leq b,c<b_1\qquad a_1<a_2\leq b,c< b_2\\
\ang{a}{n}\bracket{nb_1a_1}{c}\bracket{nb_1 a_1 a_2 b_2}{b}&\qquad a<a_1\leq b,c<b_1\qquad a_1<a_2\leq b< b_2\leq c\\
\ang{n}{a}\ang{c}{n}\bracket{nb_1a_1}{b}x_{a_1b_1}^2x_{a_2b_2}^2&\qquad a<a_1\leq b<b_1\leq c \qquad b<a_2,b_2\leq c\\
\ang{n}{a}\ang{n}{c}x_{a_1b_1}^2\bracket{nb_1a_1a_2b_2}{b}&\qquad a<a_1\leq b<b_1\leq c \qquad a_2\leq b <b_2\\
\ang{a}{b}\bracket{nb_1a_1}{n}\bracket{nb_1a_1b_2a_2}{c}&\qquad a_1\leq a,b,c<b_1\qquad b<a_2\leq c<b_2\\
\ang{c}{b}\bracket{nb_1a_1}{n}\bracket{nb_1a_1b_2a_2}{a}&\qquad a_1\leq a,b,c<b_1 \qquad a<a_2\leq b, c<b_2\\
\ang{b}{a}\bracket{nb_1a_1}{n}\bracket{nb_1a_1}{c}x_{a_2b_2}^2&\qquad a_1\leq a,b,c<b_1\qquad b<a_2,b_2\leq c\\
\bracket{nb_1a_1}{n}\begin{aligned}[t](&x_{a_2b_2}^2\bracket{nb_1a_1}{a}\ang{b}{c}\\&+\bracket{nb_1a_1a_2b_2}{b}\ang{a}{c})\end{aligned}&\qquad a_1\leq a,b,c<b_1\qquad a<a_2\leq b<b_2\leq c\\
\ang{a}{b}\bracket{nb_1a_1}{n}\bracket{nb_1a_1a_2b_2}{c}&\qquad a_1\leq a,b,c<b_1\qquad a_2\leq a, b<b_2\leq c\\
\ang{c}{b}\bracket{nb_1a_1}{n}\bracket{nb_1a_1a_2b_2}{a}&\qquad a_1\leq a,b,c<b_1\qquad a_1<a_2\leq a<b_2\leq b\\
\ang{b}{c}\bracket{nb_1a_1}{a}\bracket{nb_1a_1}{n}x_{a_2 b_2}&\qquad a_1\leq a,b,c<b_1\qquad a_1<a_2\leq a<b_2\leq b\\
\ang{c}{n}\ang{a}{b}\bracket{na_1}{nb_1}x_{a_1b_1}^2x_{a_2b_2}^2&\qquad a_1\leq a,b<b_1\leq c\qquad b<a_2,b_2\leq b_1\\
\ang{c}{n}\begin{aligned}[t](&x_{a_2b_2}^2\bracket{nb_1a_1}{a}\bracket{na_1b_1}{b}\\&+\bracket{na_1b_1}{a}\bracket{nb_1a_1a_2b_2}{b})\end{aligned}&\qquad a_1\leq a,b<b_1\leq c\qquad a<a_2\leq b<b_2\\
\ang{n}{c}\ang{a}{b}\bracket{nb_1a_1a_2b_2}{na_1 b_1}&\qquad a_1\leq a,b<b_1\leq c\qquad a_2\leq a, b<b_2\\
\ang{c}{n}\bracket{nb_1a_1}{a}\bracket{na_1b_1}{b}x_{a_2b_2}^2&\qquad a_1\leq a,b<b_1\leq c\qquad a<a_2,b_2\leq b\\
\ang{n}{c}\bracket{na_1b_1}{b}\bracket{nb_1a_1a_2b_2}{a}&\qquad a_1\leq a,b<b_1\leq c\qquad a_2\leq a<b_2\leq b
\end{cases}\label{eq:D1}
\end{equation}
for $1<a_1<a_2<b_2\leq b_1<n$ and
\begin{equation}\label{eq:D2}
D^{abc}_{\!\scriptscriptstyle 2}=\begin{cases}
\ang{n}{a}\bracket{nb_1a_1}{b}\bracket{nb_2a_2}{c}& \qquad a<a_1\leq b<b_1\leq c\qquad a_2\leq c< b_2\\
\ang{n}{a}\ang{c}{n}\bracket{nb_1a_1}{b}x_{a_2b_2}^2& \qquad a<a_1\leq b<b_1\leq c\qquad b_2\leq c\\
\ang{n}{a}\ang{b}{n}\bracket{nb_2a_2}{c}x_{a_1b_1}^2& \qquad a<a_1,b_1\leq b\qquad b<a_2\leq c<b_2\\
\ang{n}{a}\ang{b}{n}\ang{n}{c}x_{a_1b_1}^2x_{a_2b_2}^2& \qquad a<a_1,b_1\leq b\qquad b<a_2,b_2\leq c\\
\ang{n}{a}\ang{b}{c}\bracket{nb_2a_2}{n}x_{a_1b_1}^2& \qquad a<a_1,b_1\leq b\qquad a_2\leq b,c<b_2\\
\ang{n}{a}\ang{c}{n}\bracket{nb_2a_2}{b}x_{a_1b_1}^2& \qquad a<a_1,b_1\leq b\qquad a_2\leq b< b_2\leq c\\
\ang{a}{b}\bracket{na_1b_1}{n}\bracket{nb_2a_2}{c}& \qquad a_1\leq a,b<b_1\leq c\qquad a_2\leq c< b_2\\
\ang{c}{n}\ang{a}{b}\bracket{na_1b_1}{n}x_{a_2b_2}^2& \qquad a_1\leq a,b<b_1\leq c\qquad b_2\leq c\\
\ang{n}{b}\bracket{na_1b_1}{a}\bracket{nb_2a_2}{c}& \qquad a_1\leq a<b_1\leq b\qquad b< a_2\leq c< b_2\\
\ang{n}{b}\ang{c}{n}\bracket{na_1b_1}{a}x_{a_2b_2}^2& \qquad a_1\leq a<b_1\leq b\qquad b< a_2,b_2\leq c\\
\ang{c}{b}\bracket{na_1b_1}{a}\bracket{nb_2a_2}{n}& \qquad a_1\leq a<b_1\leq b\qquad a_2\leq b,c<b_2\\
\ang{n}{c}\bracket{na_1b_1}{a}\bracket{na_2b_2}{b}& \qquad a_1\leq a<b_1\leq b\qquad a_2\leq b<b_2\leq c
\end{cases}
\end{equation}
for $1<a_1<b_1\leq a_2<b_2<n$.  For other orderings of $a,b,c$ one can
use the total antisymmetry of $D_1^{abc}$ and $D_2^{abc}$ under permutations
of $a,b,c$.
It is quite astonishing that in 28 out of 30 cases these determinants
are given by a single term.  
This formula is implemented in {\tt GGT} by {\tt GGTnnmhvgluon}.


\section{Explicit formulae for trees with fermions}
\label{explicit-fermions}

Here we explicitly write out our formulas~(\ref{masterquark}) and \eqref{master_quark_SYM_original} for the
MHV, NMHV and NNMHV cases with up to six fermions.
\label{explicit2}
\subsection{MHV amplitudes}
The simplest amplitudes involving fermions are the MHV amplitudes. The
amplitudes with one negative-helicity gluon $a$ and two fermions of
opposite helicity and the same flavor, 
$b$ ($+\frac12$) and $\bar{c}$ ($-\frac12$) (or vice-versa), 
are given by
\begin{align}
 A_n(a^-, b,\bar{c})
&= \delta^{(4)}(p)\frac{\ang{a}{c}^3\ang{a}{b}}
{\ang{1}{2}\dots\ang{n}{1}} \,, \\
A_n(a^-, \bar{b},c)
&=-\delta^{(4)}(p)\frac{\ang{a}{b}^3\ang{a}{c}}
{\ang{1}{2}\dots\ang{n}{1}}\,.
\end{align}
These formulae correspond to case (1) in figure~\ref{fig-fermions12}.
We note that the latter formula is related to the former one by a
reflection symmetry, under which the cyclic ordering is reversed
and there is a relabeling $b \leftrightarrow c$.
In the NMHV case we will omit formulae that can be obtained from the
presented formulae by a reflection symmetry.

An equally compact formula can be obtained for the MHV amplitudes with
four fermions and only positive-helicity gluons:
\begin{align} 
 A_n(a^A, \bar{b}^B,c^C,\bar{d}^D)
&=\frac{\delta^{(4)}(p)\ang{b}{d}^2}{\ang{1}{2}\dots \ang{n}{1}}
\left(\delta^{AB}\delta^{CD}\ang{d}{a}\ang{c}{b}
-\delta^{AD}\delta^{BC}\ang{d}{c}\ang{a}{b}\right) \,,
\label{abbcddMHV} \\
 A_n(a^A, b^B,\bar{c}^C,\bar{d}^D)
&=\frac{\delta^{(4)}(p)\ang{c}{d}^2}{\ang{1}{2}\dots \ang{n}{1}}
\left(\delta^{AD}\delta^{BC}\ang{d}{b}\ang{a}{c}
-\delta^{AC}\delta^{BD}\ang{d}{a}\ang{b}{c}\right)\,,
\label{abccddMHV}
\end{align}
which in the single-flavor case simplifies to
\begin{align}
 A_n(a, \bar{b},c,\bar{d})
&=\frac{\delta^{(4)}(p)\ang{b}{d}^3\ang{a}{c}}{\ang{1}{2}\dots \ang{n}{1}}
\label{abbcddMHVsingle} \,, \\
 A_n(a, b,\bar{c},\bar{d})
&=-\frac{\delta^{(4)}(p)\ang{c}{d}^3\ang{a}{b}}{\ang{1}{2}\dots \ang{n}{1}}\,.
\label{abccddMHVsingle}
\end{align}
\Eqn{abccddMHVsingle} corresponds to case (2a) in
figure~\ref{fig-fermions12}, whereas \eqn{abbcddMHV}
for $A = B \neq C = D$ corresponds to case (2b).

To complete the list of MHV amplitudes with up to four fermions we
also give the MHV amplitude with four positive-helicity fermions and one
negative-helicity gluon:
\begin{align}
A_n(a^A, b^B,c^C,d^D,n^-)&=
\int d\eta^A_a\int d\eta^B_b\int d\eta_c^C \int d\eta_d^D
\int d^4\eta_n \, {\mathcal A}_n^{\mathrm MHV}\notag\\
&=\frac{\delta^{(4)}(p)\epsilon^{ABCD}}
{\ang{1}{2}\dots \ang{n}{1}}\ang{n}{a}\ang{n}{b}\ang{n}{c}\ang{n}{d}\,.
\label{fourpf}
\end{align}
This amplitude is not needed for QCD.

\subsection{NMHV amplitudes}
\subsubsection{Two fermions}
To illustrate the use of our master formula~\eqref{masterquark} we
compute the NMHV amplitude with two opposite-helicity fermions at
positions $a$, $\bar{b}$ and two negative-helicity gluons at positions
$c$ and $n$. At this stage we leave the color order
arbitrary. Starting with the path-matrix
\begin{equation}
\Xi^{\text{path}}=\left (
\begin{array}{llllll}
\ang{n}{c}  & \ang{n}{a}&\ang{n}{\bar{b}} \cr
(\Xi_{n})^{c}_{st} & (\Xi_{n})^{a}_{st}& (\Xi_{n})^{\bar{b}}_{st}  \cr
\end{array}\right )
\end{equation}
we can immediately write down the amplitude
\begin{equation}
(A_n)^{\text{NMHV}}_{q\bar{q}}
=\frac{\delta^{(4)}(p)\sign(\tau)}{\ang{1}{2}\dots\ang{n}{1}}
\sum_{1<s<t<n}\tilde{R}_{n;st}D_{n;st}^{ca}
\left(D_{n;st}^{c\bar{b}}\right)^3\,,
\label{eq:NMHV2fermion}
\end{equation}
where the $2\times 2$ determinant $D_{n;st}^{ab}$ has been defined
in \eqn{eq:Dsymbol}. As already stated, the last equation holds for an
arbitrary color ordering.
In the following we take $a<b<c$ and specify the color ordering:
\begin{align}
A_n(a,b^-,\bar{c},n^-)&=\frac{\delta^{(4)}(p)}
{\ang{1}{2}\dots\ang{n}{1}}\biggl[
{}-\ang{a}{b}\ang{b}{c}^3\sum_{1<s\leq a,b,c<t< n}
\bracket{nts}{n}^4\tilde{R}_{n,st}\notag\\
&\=\phantom{\frac{\delta^{(4)}(p)}
{\ang{1}{2}\dots\ang{n}{1}}\biggl[}
-\ang{b}{c}^3\ang{a}{n}\sum_{a<s\leq b,c<t<n}
\bracket{nts}{b}\bracket{nts}{n}^3\tilde{R}_{n,st}\notag\\
&\=\phantom{\frac{\delta^{(4)}(p)}
{\ang{1}{2}\dots\ang{n}{1}}\biggl[}
-\ang{c}{n}^3\ang{a}{n}\sum_{a<s\leq b<t\leq c}
\bracket{nst}{b}^3\bracket{nts}{b}\tilde{R}_{n,st}\notag\\
&\=\phantom{\frac{\delta^{(4)}(p)}
{\ang{1}{2}\dots\ang{n}{1}}\biggl[}
-\ang{c}{n}^3\ang{a}{b}\sum_{1<s\leq a,b<t\leq c}
\bracket{nst}{b}^3\bracket{nts}{n}\tilde{R}_{n,st}\biggr]\,,
\end{align}
\begin{align}
A_n(a,\bar{b},c^-,n^-)&=\frac{\delta^{(4)}(p)}
{\ang{1}{2}\dots\ang{n}{1}}\biggl[{}
+\ang{a}{c}\ang{b}{c}^3\sum_{1<s\leq a,b,c<t< n}
\bracket{nts}{n}^4\tilde{R}_{n,st}\notag\\
&\=\phantom{\frac{\delta^{(4)}(p)}
{\ang{1}{2}\dots\ang{n}{1}}\biggl[}
+\ang{a}{n}\ang{b}{n}^3\sum_{b<s\leq c<t<n}
\bracket{nts}{c}^4\tilde{R}_{n,st}\notag\\
&\=\phantom{\frac{\delta^{(4)}(p)}
{\ang{1}{2}\dots\ang{n}{1}}\biggl[}
+\ang{n}{c}^4\ang{a}{n}\ang{b}{n}^3\sum_{b<s<t\leq c }
(x_{st}^2)^4\tilde{R}_{n,st}\notag\\
&\=\phantom{\frac{\delta^{(4)}(p)}
{\ang{1}{2}\dots\ang{n}{1}}\biggl[}
+\ang{c}{n}^4\sum_{1<s\leq a,b<t\leq c}
\bracket{nst}{b}^3\bracket{nst}{a}\tilde{R}_{n,st}\notag\\
&\=\phantom{\frac{\delta^{(4)}(p)}
{\ang{1}{2}\dots\ang{n}{1}}\biggl[}
+\ang{b}{c}^3\ang{a}{n}\sum_{a<s\leq b,c<t<n}
\bracket{nts}{c}\bracket{nts}{n}^3\tilde{R}_{n,st}\notag\\
&\=\phantom{\frac{\delta^{(4)}(p)}
{\ang{1}{2}\dots\ang{n}{1}}\biggl[}
+\ang{c}{n}^4\ang{a}{n}\sum_{a<s\leq b<t\leq c}
x_{st}^2\bracket{nst}{b}^3\tilde{R}_{n,st}\biggr]\,.
\end{align}
These simplified expressions are implemented in {\tt GGT} by 
{\tt GGTnmhv2ferm} (see appendix \ref{GGT}).

\subsubsection{Four Fermions}
We proceed with the NMHV amplitude with four fermions at positions
$a_1^{A_1},a_2^{A_2},\bar{b}_1^{B_1},\bar{b}_2^{B_2}$ and one
negative-helicity gluon. Without loss of generality we put the
negative-helicity gluon at position $n$. Again we leave the color
ordering arbitrary. A straightforward application of our formulas
\eqref{master_quark_SYM_original} and \eqref{masterquark} yields
\begin{equation}
\left(A_n\right)^{\text{NMHV}}_{(q\bar{q})^2}=\frac{\delta^{(4)}(p)\sign(\tau)}{\ang{1}{2}\dots \ang{n}{1}}\sum_{1<s<t<n}\!\!\tilde{R}_{n;st}\left(D_{n;st}^{\bar{b}_1\bar{b}_2}\right)^2\left(\delta^{\scriptscriptstyle A_1B_1}\delta^{\scriptscriptstyle A_2B_2} D_{n;st}^{a_1\bar{b}_2}D_{n;st}^{\bar{b}_1a_2}-\delta^{\scriptscriptstyle \scriptscriptstyle A_1B_2}\delta^{\scriptscriptstyle A_2B_1} D_{n;st}^{a_2\bar{b}_2}D_{n;st}^{\bar{b}_1a_1}\right)
\label{eq:NMHV4fermion}
\end{equation}
in the ${\cal N}=4$ super Yang-Mills case, and
\begin{equation}
\left(A_n\right)^{\text{NMHV}}_{(q\bar{q})^2}
=\frac{\delta^{(4)}(p)\sign(\tau)}
{\ang{1}{2}\dots \ang{n}{1}}\sum_{1<s<t<n}
\tilde{R}_{n;st}\left(D_{n;st}^{\bar{b}_1\bar{b}_2}\right)^3D_{n;st}^{a_1a_2}
\label{eq:NMHV4fermionS}
\end{equation}
for single-flavor QCD, with $D^{ab}_{n;st}$ defined in
equation~\eqref{eq:Dsymbol}. Taking $a<b<c<d$ we now specify the
color ordering,
\begin{align}
A_n(a^A,b^B,\bar{c}^C,&\bar{d}^D,n^-)=\frac{\delta^{(4)}(p)}{\ang{1}{2}\dots \ang{n}{1}}\times\notag\\
&\=\times\biggl[{}+\delta^{AC}\delta^{BD}\ang{a}{n}\ang{b}{c}\ang{c}{d}^2\sum_{a<s\leq b,d<t<n}\bracket{nts}{d}\bracket{nts}{n}^3\tilde{R}_{n,st}-(c\leftrightarrow d)\notag\\
&\=\phantom{\times\biggl[}+\delta^{AC}\delta^{BD}\ang{a}{d}\ang{b}{c}\ang{d}{c}^2\sum_{1<s\leq a,d<t<n}\bracket{nts}{n}^4\tilde{R}_{n,st}-(a\leftrightarrow b)\notag\\
&\=\phantom{\times\biggl[}+\delta^{AC}\delta^{BD}\ang{n}{a}\ang{b}{c}\ang{d}{n}^3\sum_{a<s\leq b,c<t\leq d}x_{st}^2\bracket{nst}{n}\bracket{nst}{c}^2\tilde{R}_{n,st}\notag\\
&\=\phantom{\times\biggl[++}-\delta^{AD}\delta^{BC}\ang{a}{n}\ang{d}{n}^3\sum_{a<s\leq b,c<t\leq d}\bracket{nts}{c}\bracket{nst}{b}\bracket{nst}{c}^2\tilde{R}_{n,st}\notag\\
&\=\phantom{\times\biggl[}+\delta^{AC}\delta^{BD}\ang{b}{c}\ang{d}{n}^3\sum_{1<s\leq a,c<t\leq d}\bracket{nst}{a}\bracket{nts}{n}\bracket{nst}{c}^2\tilde{R}_{n,st}-(a \leftrightarrow b)\biggr]\,, 
\end{align}
where ``$(c\leftrightarrow d)$'' implies the substitution $c\leftrightarrow d$
in the arguments of the spinor strings, as well as the corresponding
substitution $C\leftrightarrow D$ in the arguments of the $\delta$ functions,
but {\it no} change in the summation range.
The other inequivalent orderings of quarks and anti-quarks are,
\begin{align}
 A_n(a^A,\bar{b}^B,c^C,&\bar{d}^D,n^-)=\frac{\delta^{(4)}(p)}{\ang{1}{2}\dots \ang{n}{1}}\times\notag\\
&\=\times\biggl[{}+\delta^{AB}\delta^{CD}\ang{a}{n}\ang{b}{n}^3\sum_{b<s\leq c,d<t<n}\bracket{nts}{c}\bracket{nts}{d}^3\tilde{R}_{n,st}\notag\\
&\=\phantom{\times\biggl[}+\delta^{AB}\delta^{CD}\ang{n}{a}\ang{c}{b}\ang{b}{d}^2\sum_{a<s\leq b,d<t<n}\bracket{nts}{d}\bracket{nts}{n}^3\tilde{R}_{n,st}-(b\leftrightarrow d)\notag\\
&\=\phantom{\times\biggl[}+\delta^{AB}\delta^{CD}\ang{a}{d}\ang{b}{c}\ang{d}{b}^2\sum_{1<s\leq a,d<t<n}\bracket{nts}{n}^4\tilde{R}_{n,st}-(a\leftrightarrow c)\notag\\
&\=\phantom{\times\biggl[}+\delta^{AB}\delta^{CD}\ang{n}{a}\ang{d}{n}^3\ang{n}{b}^3\sum_{b<s\leq c<t\leq d}\bracket{nts}{c}
(x_{st}^2)^3\tilde{R}_{n,st}\notag\\
&\=\phantom{\times\biggl[}+\delta^{AB}\delta^{CD}\ang{n}{a}\ang{n}{c}\ang{n}{b}^3\ang{n}{d}^3\sum_{b<s<t\leq c}
(x_{st}^2)^4\tilde{R}_{n,st}\notag\\
&\=\phantom{\times\biggl[}+\delta^{AB}\delta^{CD}\ang{n}{a}\ang{b}{c}\ang{d}{n}^3\sum_{a<s\leq b,c<t\leq d}x_{st}^2\bracket{nst}{n}\bracket{nst}{b}^2\tilde{R}_{n,st}\notag\\
&\=\phantom{\times\biggl[++}-\delta^{AD}\delta^{CB}\ang{n}{a}\ang{d}{n}^3\sum_{a<s\leq b,c<t\leq d}\bracket{nts}{b}\bracket{nst}{c}\bracket{nst}{b}^2\tilde{R}_{n,st}\notag\\
&\=\phantom{\times\biggl[}+\delta^{AB}\delta^{CD}\ang{n}{a}\ang{n}{c}\ang{d}{n}^3\sum_{a<s\leq b<t\leq c}x_{st}^2\bracket{nst}{b}^3\tilde{R}_{n,st}\notag\\
&\=\phantom{\times\biggl[}+\delta^{AB}\delta^{CD}\ang{b}{c}\ang{d}{n}^3\sum_{1<s\leq a,c<t\leq d}\bracket{nst}{a}\bracket{nts}{n}\bracket{nst}{b}^2\tilde{R}_{n,st}-(a\ \leftrightarrow c)\notag\\
&\=\phantom{\times\biggl[}+\delta^{AB}\delta^{CD}\ang{n}{c}\ang{n}{d}^3\sum_{1<s\leq a,b<t\leq c}\bracket{nst}{a}\bracket{nst}{b}^3\tilde{R}_{n,st}\biggr]\,,
\end{align}
\begin{align}
 A_n(a^A,\bar{b}^B,&\bar{c}^C,d^D,n^-)=-\frac{\delta^{(4)}(p)}{\ang{1}{2}\dots \ang{n}{1}}\times\notag\\
&\=\times\biggl[{}+\delta^{AB}\delta^{CD}\ang{a}{n}\ang{b}{n}^3\sum_{b<s\leq c,d<t<n}\bracket{nts}{d}\bracket{nts}{c}^3\tilde{R}_{n,st}\notag\\
&\=\phantom{\times\biggl[}+\delta^{AB}\delta^{CD}\ang{n}{a}\ang{n}{d}\ang{b}{n}^3\sum_{b<s\leq c<t\leq d}x_{st}^2\bracket{nts}{c}^3\tilde{R}_{n,st}\notag\\
&\=\phantom{\times\biggl[}+\delta^{AB}\delta^{CD}\ang{n}{a}\ang{d}{b}\ang{b}{c}^2\sum_{a<s\leq b,d<t<n}\bracket{nts}{c}\bracket{nts}{n}^3\tilde{R}_{n,st}-(b\leftrightarrow c)\notag\\
&\=\phantom{\times\biggl[}+\delta^{AB}\delta^{CD}\ang{n}{a}\ang{n}{d}\ang{b}{c}^2\sum_{a<s\leq b,c<t\leq d}\bracket{nts}{c}\bracket{nst}{b}\bracket{nts}{n}^2\tilde{R}_{n,st}-(b\leftrightarrow c)\notag\\
&\=\phantom{\times\biggl[}+\delta^{AB}\delta^{CD}\ang{a}{c}\ang{b}{d}\ang{c}{b}^2\sum_{1<s\leq a,d<t<n}\bracket{nts}{n}^4\tilde{R}_{n,st}-(a\leftrightarrow d)\notag\\
&\=\phantom{\times\biggl[}+\delta^{AB}\delta^{CD}\ang{a}{c}\ang{n}{d}\ang{c}{b}^2\sum_{1<s\leq a,c<t\leq d}\bracket{nst}{b}\bracket{nst}{n}^3\tilde{R}_{n,st}-(b\leftrightarrow c)\notag\\
&\=\phantom{\times\biggl[}+\delta^{AB}\delta^{CD}\ang{n}{a}\ang{n}{d}\ang{n}{b}^3\ang{n}{c}^3\sum_{b<s<t\leq c}(x_{st}^2)^4\tilde{R}_{n,st}\notag\\
&\=\phantom{\times\biggl[}+\delta^{AB}\delta^{CD}\ang{n}{a}\ang{n}{d}\ang{c}{n}^3\sum_{a<s\leq b<t\leq c}x_{st}^2\bracket{nst}{b}^3\tilde{R}_{n,st}\notag\\
&\=\phantom{\times\biggl[}+\delta^{AB}\delta^{CD}\ang{n}{d}\ang{n}{c}^3\sum_{1<s\leq a,b<t\leq c}\bracket{nst}{a}\bracket{nst}{b}^3\tilde{R}_{n,st}\biggr]\,,
\end{align}
\begin{align}
A_n(\bar{a}^A,b^B&,c^C,\bar{d}^D,n^-)=-\frac{\delta^{(4)}(p)}{\ang{1}{2}\dots \ang{n}{1}}\times\notag\\
&\=\times\biggl[{}+\delta^{AB}\delta^{CD}\ang{b}{n}\ang{a}{n}^3\sum_{b<s\leq c,d<t<n}\bracket{nts}{c}\bracket{nts}{d}^3\tilde{R}_{n,st}\notag\\
&\=\phantom{\times\biggl[}+\delta^{AB}\delta^{CD}\ang{b}{d}\ang{a}{n}^3\sum_{a<s\leq b,d<t<n}\bracket{nts}{n}\bracket{nts}{c}\bracket{nts}{d}^2\tilde{R}_{n,st}-(b\leftrightarrow c)\notag\\
&\=\phantom{\times\biggl[}+\delta^{AB}\delta^{CD}\ang{b}{d}\ang{a}{c}\ang{d}{a}^2\sum_{1<s\leq a,d<t<n}\bracket{nts}{n}^4\tilde{R}_{n,st}-(b\leftrightarrow c)\notag\\
&\=\phantom{\times\biggl[}+\delta^{AB}\delta^{CD}\ang{n}{b}\ang{d}{n}^3\ang{n}{a}^3\sum_{b<s\leq c<t\leq d}\bracket{nts}{c}(x_{st}^2)^3\tilde{R}_{n,st}\notag\\
&\=\phantom{\times\biggl[}+\delta^{AB}\delta^{CD}\ang{n}{b}\ang{n}{c}\ang{n}{a}^3\ang{n}{d}^3\sum_{b<s<t\leq c}(x_{st}^2)^4\tilde{R}_{n,st}\notag\\
&\=\phantom{\times\biggl[}+\delta^{AB}\delta^{CD}\ang{n}{a}^3\ang{n}{d}^3\sum_{a<s\leq b,c<t\leq d}\bracket{nst}{b}\bracket{nts}{c}(x_{st}^2)^2\tilde{R}_{n,st}-(b\ \leftrightarrow c)\notag\\
&\=\phantom{\times\biggl[}+\delta^{AB}\delta^{CD}\ang{n}{c}\ang{d}{n}^3\ang{n}{a}^3\sum_{a<s\leq b<t\leq c}\bracket{nst}{b}(x_{st}^2)^3\tilde{R}_{n,st}\notag\\
&\=\phantom{\times\biggl[}+\delta^{AB}\delta^{CD}\ang{a}{c}\ang{d}{n}^3\sum_{1<s\leq a,c<t\leq d}\bracket{nst}{b}\bracket{nts}{n}\bracket{nst}{a}^2\tilde{R}_{n,st}-(b\ \leftrightarrow c)\notag\\
&\=\phantom{\times\biggl[}+\delta^{AB}\delta^{CD}\ang{n}{c}\ang{n}{d}^3\sum_{1<s\leq a,b<t\leq c}\bracket{nst}{b}\bracket{nst}{a}^3\tilde{R}_{n,st}\biggr]\,.
\end{align}
For $A\neq B$, and all fermions cyclically adjacent we have 
\begin{multline}
A_n(a^A,\overline{(a\+1)}^B,(a\+2)^B,\overline{(a\+3)}^A,n^-)=\frac{\delta^{(4)}(p)}{\ang{1}{2}\dots \ang{n}{1}}\times\\
\begin{aligned}
 \biggl[&\ang{a}{n}\ang{a\+2}{a\+3}\ang{a\+1}{a\+3}^2\sum_{a\+3<t< n}\bracket{nta\+2}{a\+1}\bracket{nta\+1}{n}^3\tilde{R}_{n,a\+1t}\\
&+\ang{a\+2}{a\+3}\ang{a}{a\+1}\ang{a\+1}{a\+3}^2\sum_{1<s\leq a,a\+3<t<n}\bracket{nts}{n}^4\tilde{R}_{n,st}\\
&+\ang{a}{n}\ang{a\+3}{n}^3\ang{a\+1}{a\+2}^4\bra{n}x_{n\,a\+3}\cket{a\+2}\bra{n}x_{n\,a\+1}\cket{a\+1}\bra{n}x_{n\,a\+1}\cket{a\+2}^2\tilde{R}_{n,a\+1a\+3}\\
&-\ang{a}{a\+1}\ang{a\+3}{n}^3\sum_{1<s\leq a}\bracket{nsa\+3}{a\+2}\bracket{nsa\+3}{n}\bracket{nsa\+3}{a\+1}^2\tilde{R}_{n,sa\+3}\biggr] \,.
\end{aligned}
\end{multline}
This amplitude may be used to generate the NMHV amplitudes for 
$V q\bar{q} g \ldots g$, as discussed in section 3.

In the single-flavor case we obtain
\begin{align}
 A_n(a,b,\bar{c},\bar{d},n^-)&=\frac{\delta^{(4)}(p)}{\ang{1}{2}\dots \ang{n}{1}}\biggl[{}+\ang{n}{a}\ang{c}{d}^3\sum_{a<s\leq b,d<t<n}\bracket{nts}{b}\bracket{nts}{n}^3\tilde{R}_{n,st}\notag\\
&\=\phantom{\frac{\delta^{(4)}(p)}{\ang{1}{2}\dots \ang{n}{1}}\biggl[}+\ang{a}{b}\ang{d}{c}^3\sum_{1<s\leq a,d<t<n}\bracket{nts}{n}^4\tilde{R}_{n,st}\notag\\
&\=\phantom{\frac{\delta^{(4)}(p)}{\ang{1}{2}\dots \ang{n}{1}}\biggl[}+\ang{n}{a}\ang{d}{n}^3\sum_{a<s\leq b,c<t\leq d}\bracket{nts}{b}\bracket{nst}{c}^3\tilde{R}_{n,st}\notag\\
&\=\phantom{\frac{\delta^{(4)}(p)}{\ang{1}{2}\dots \ang{n}{1}}\biggl[}+\ang{b}{a}\ang{d}{n}^3\sum_{1<s\leq a,c<t\leq d}
\bracket{nts}{n}\bracket{nst}{c}^3\tilde{R}_{n,st}\biggr] \,,
\end{align}
\begin{align}
 A_n(a,\bar{b},c,\bar{d},n^-)
&=\frac{\delta^{(4)}(p)}{\ang{1}{2}\dots \ang{n}{1}}\biggl[{}
+\ang{a}{n}\ang{b}{n}^3\sum_{b<s\leq c,d<t<n}
\bracket{nts}{c}\bracket{nts}{d}^3\tilde{R}_{n,st}\notag\\
&\=\phantom{\frac{\delta^{(4)}(p)}{\ang{1}{2}\dots \ang{n}{1}}\biggl[}
+\ang{n}{a}\ang{d}{b}^3\sum_{a<s\leq b,d<t<n}
\bracket{nts}{c}\bracket{nts}{n}^3\tilde{R}_{n,st}\notag\\
&\=\phantom{\frac{\delta^{(4)}(p)}{\ang{1}{2}\dots \ang{n}{1}}\biggl[}
+\ang{a}{c}\ang{b}{d}^3\sum_{1<s\leq a,d<t<n}
\bracket{nts}{n}^4\tilde{R}_{n,st}\notag\\
&\=\phantom{\frac{\delta^{(4)}(p)}{\ang{1}{2}\dots \ang{n}{1}}\biggl[}
+\ang{n}{a}\ang{d}{n}^3\ang{n}{b}^3
\sum_{b<s\leq c<t\leq d}\bracket{nts}{c}
(x_{st}^2)^3\tilde{R}_{n,st}\notag\\
&\=\phantom{\frac{\delta^{(4)}(p)}{\ang{1}{2}\dots \ang{n}{1}}\biggl[}
+\ang{n}{a}\ang{n}{c}\ang{n}{b}^3\ang{n}{d}^3
\sum_{b<s<t\leq c}(x_{st}^2)^4\tilde{R}_{n,st}\notag\\
&\=\phantom{\frac{\delta^{(4)}(p)}{\ang{1}{2}\dots \ang{n}{1}}\biggl[}
+\ang{n}{a}\ang{n}{d}^3
\sum_{a<s\leq b,c<t\leq d}
\bracket{nts}{c}\bracket{nst}{b}^3\tilde{R}_{n,st}\notag\\
&\=\phantom{\frac{\delta^{(4)}(p)}{\ang{1}{2}\dots \ang{n}{1}}\biggl[}
+\ang{n}{a}\ang{n}{c}\ang{d}{n}^3\sum_{a<s\leq b<t\leq c}
x_{st}^2\bracket{nst}{b}^3\tilde{R}_{n,st}\notag\\
&\=\phantom{\frac{\delta^{(4)}(p)}{\ang{1}{2}\dots \ang{n}{1}}\biggl[}
+\ang{a}{c}\ang{d}{n}^3
\sum_{1<s\leq a,c<t\leq d}
\bracket{nts}{n}\bracket{nst}{b}^3\tilde{R}_{n,st}\notag\\
&\=\phantom{\frac{\delta^{(4)}(p)}{\ang{1}{2}\dots \ang{n}{1}}\biggl[}
+\ang{n}{c}\ang{n}{d}^3
\sum_{1<s\leq a,b<t\leq c}
\bracket{nst}{a}\bracket{nst}{b}^3\tilde{R}_{n,st}\biggr] \,,
\end{align}
\begin{align}
 A_n(a,\bar{b},\bar{c},d,n^-)&=
-\frac{\delta^{(4)}(p)}{\ang{1}{2}\dots \ang{n}{1}}\biggl[{}+\ang{a}{n}\ang{b}{n}^3
\sum_{b<s\leq c,d<t<n}
\bracket{nts}{d}\bracket{nts}{c}^3\tilde{R}_{n,st}\notag\\
&\=\phantom{-\frac{\delta^{(4)}(p)}{\ang{1}{2}\dots \ang{n}{1}}\biggl[}+\ang{n}{a}\ang{n}{d}\ang{b}{n}^3
\sum_{b<s\leq c<t\leq d}x_{st}^2\bracket{nts}{c}^3\tilde{R}_{n,st}\notag\\
&\=\phantom{-\frac{\delta^{(4)}(p)}{\ang{1}{2}\dots \ang{n}{1}}\biggl[}+\ang{n}{a}\ang{c}{b}^3\sum_{a<s\leq b,d<t<n}
\bracket{nts}{d}\bracket{nts}{n}^3\tilde{R}_{n,st}\notag\\
&\=\phantom{-\frac{\delta^{(4)}(p)}{\ang{1}{2}\dots \ang{n}{1}}\biggl[}+\ang{n}{a}\ang{n}{d}\ang{b}{c}^3
\sum_{a<s\leq b,c<t\leq d}
x_{st}^2 \bracket{nts}{n}^3
\tilde{R}_{n,st}\notag\\
&\=\phantom{-\frac{\delta^{(4)}(p)}{\ang{1}{2}\dots \ang{n}{1}}\biggl[}+\ang{a}{d}\ang{b}{c}^3
\sum_{1<s\leq a,d<t<n}\bracket{nts}{n}^4\tilde{R}_{n,st}\notag\\
&\=\phantom{-\frac{\delta^{(4)}(p)}{\ang{1}{2}\dots \ang{n}{1}}\biggl[}+\ang{b}{c}^3\ang{n}{d}
\sum_{1<s\leq a,c<t\leq d}
\bracket{nst}{a}\bracket{nst}{n}^3\tilde{R}_{n,st}\notag\\
&\=\phantom{-\frac{\delta^{(4)}(p)}{\ang{1}{2}\dots \ang{n}{1}}\biggl[}+\ang{n}{a}\ang{n}{d}\ang{n}{b}^3\ang{n}{c}^3
\sum_{b<s<t\leq c}(x_{st}^2)^4\tilde{R}_{n,st}\notag\\
&\=\phantom{-\frac{\delta^{(4)}(p)}{\ang{1}{2}\dots \ang{n}{1}}\biggl[}+\ang{n}{a}\ang{n}{d}\ang{c}{n}^3
\sum_{a<s\leq b<t\leq c}x_{st}^2\bracket{nst}{b}^3\tilde{R}_{n,st}\notag\\
&\=\phantom{-\frac{\delta^{(4)}(p)}{\ang{1}{2}\dots \ang{n}{1}}\biggl[}+\ang{n}{d}\ang{n}{c}^3
\sum_{1<s\leq a,b<t\leq c}
\bracket{nst}{a}\bracket{nst}{b}^3\tilde{R}_{n,st}\biggr] \,,
\end{align}
\begin{align}
A_n(\bar{a},b,c,\bar{d},n^-)&
=-\frac{\delta^{(4)}(p)}{\ang{1}{2}\dots \ang{n}{1}}\biggl[{}+\ang{b}{n}\ang{a}{n}^3\sum_{b<s\leq c,d<t<n}
\bracket{nts}{c}\bracket{nts}{d}^3\tilde{R}_{n,st}\notag\\
&\=\phantom{-\frac{\delta^{(4)}(p)}{\ang{1}{2}\dots \ang{n}{1}}\biggl[}+\ang{b}{c}\ang{a}{n}^3\sum_{a<s\leq b,d<t<n}
\bracket{nts}{n}\bracket{nts}{d}^3\tilde{R}_{n,st}\notag\\
&\=\phantom{-\frac{\delta^{(4)}(p)}{\ang{1}{2}\dots \ang{n}{1}}\biggl[}+\ang{b}{c}\ang{a}{d}^3
\sum_{1<s\leq a,d<t<n}\bracket{nts}{n}^4\tilde{R}_{n,st}\notag\\
&\=\phantom{-\frac{\delta^{(4)}(p)}{\ang{1}{2}\dots \ang{n}{1}}\biggl[}+\ang{n}{b}\ang{d}{n}^3\ang{n}{a}^3
\sum_{b<s\leq c<t\leq d}\bracket{nts}{c}(x_{st}^2)^3\tilde{R}_{n,st}\notag\\
&\=\phantom{-\frac{\delta^{(4)}(p)}{\ang{1}{2}\dots \ang{n}{1}}\biggl[}+\ang{n}{b}\ang{n}{c}\ang{n}{a}^3\ang{n}{d}^3
\sum_{b<s<t\leq c}(x_{st}^2)^4\tilde{R}_{n,st}\notag\\
&\=\phantom{-\frac{\delta^{(4)}(p)}{\ang{1}{2}\dots \ang{n}{1}}\biggl[}+\ang{n}{a}^3\ang{n}{d}^3\ang{b}{c}
\sum_{a<s\leq b,c<t\leq d}
\bracket{nts}{n}(x_{st}^2)^3\tilde{R}_{n,st}\notag\\
&\=\phantom{-\frac{\delta^{(4)}(p)}{\ang{1}{2}\dots \ang{n}{1}}\biggl[}+\ang{n}{c}\ang{d}{n}^3\ang{n}{a}^3
\sum_{a<s\leq b<t\leq c}\bracket{nst}{b}(x_{st}^2)^3\tilde{R}_{n,st}\notag\\
&\=\phantom{-\frac{\delta^{(4)}(p)}{\ang{1}{2}\dots \ang{n}{1}}\biggl[}+\ang{b}{c}\ang{d}{n}^3
\sum_{1<s\leq a,c<t\leq d}
\bracket{nts}{n}\bracket{nst}{a}^3\tilde{R}_{n,st}\notag\\
&\=\phantom{-\frac{\delta^{(4)}(p)}{\ang{1}{2}\dots \ang{n}{1}}\biggl[}+\ang{n}{c}\ang{n}{d}^3
\sum_{1<s\leq a,b<t\leq c}
\bracket{nst}{b}\bracket{nst}{a}^3\tilde{R}_{n,st}\biggr] \,.
\end{align}
These simplified expressions are implemented in {\tt GGT} by
{\tt GGTnmhv4fermS} for the single-flavor case, and by {\tt GGTnmhv4ferm} for
the general-flavor case.  See appendix \ref{GGT} for the documentation.

\subsubsection{Six fermions}
\label{NMHVsixfermions}
In the case of the six-fermion NMHV amplitude there is no
negative-helicity gluon for us to put at position $n$ as we did
in the previous examples.  The fermions are at positions
$a_1^{A_1},a_2^{A_2},a_3^{A_3},\bar{b}_1^{B_1},\bar{b}_2^{B_2}$
and $\bar{n}^{B_3}$. 
This time the path-matrix~\eqref{Pathmatrix2} is given by
\begin{equation}
\Xi^{\text{path}}=\left (
\begin{array}{llllll}
\frac{\ang{\bar{b}_2}{\bar{b}_1}}{\ang{\bar{b}_2}{n}}&0&1&\frac{\ang{\bar{b}_2}{a_1}}{\ang{\bar{b}_2}{n}}&\frac{\ang{\bar{b}_2}{a_2}}{\ang{\bar{b}_2}{n}}&\frac{\ang{\bar{b}_2}{a_3}}{\ang{\bar{b}_2}{n}}\cr
\ang{n}{\bar{b}_1}  & \ang{n}{\bar{b}_2}&0&\ang{n}{a_1}&\ang{n}{a_2}&\ang{n}{a_3} \cr
(\Xi_{n})^{\bar{b}_1}_{st} & (\Xi_{n})^{\bar{b}_2}_{st}&0& (\Xi_{n})^{a_{1}}_{st}& (\Xi_{n})^{a_{2}}_{st}& (\Xi_{n})^{a_{3}}_{st}  \cr
\end{array}\right )\,.
\end{equation}
As ingredients of formula~\eqref{master_quark_SYM_original} we need the determinants
\begin{align}
\det\left(\Xi^{\text{path}}\rvert_q\right)&=D_{n;st}^{\bar{b}_1\bar{b}_2}\,,&
\det\left(\Xi^{\text{path}}\rvert_q(\bar{b}_{1}\rightarrow a_i)\right)&=D_{n;st}^{a_i\bar{b}_2}\,,\\
 \det\left(\Xi^{\text{path}}\rvert_q(\bar{b}_{2}\rightarrow a_i)\right)&=D_{n;st}^{\bar{b}_1a_i}\,,&\det\left(\Xi^{\text{path}}\rvert_q(\bar{n}\rightarrow a_i)\right)&=D_{n;st}^{\bar{b}_1\bar{b}_2a_i}\,.
\end{align}
We recall that $D_{n;st}^{ab}$ has been defined in eq.~\eqref{eq:Dsymbol} and the $3\times 3$ determinant $D_{n;st}^{abc}$ reads
\begin{equation}
D_{n;st}^{abc}:=\left |
\begin{array}{llllll}
\frac{\ang{b}{a}}{\ang{b}{n}}&0&\frac{\ang{b}{c}}{\ang{b}{n}}\cr
\ang{n}{a}  & \ang{n}{b}&\ang{n}{c} \cr
(\Xi_{n})^{a}_{st} & (\Xi_{n})^{b}_{st}& (\Xi_{n})^{c}_{st}\cr
\end{array}\right |=\ang{a}{b}(\Xi_{n})^{c}_{st}+\ang{b}{c}(\Xi_{n})^{a}_{st}+\ang{c}{a}(\Xi_{n})^{b}_{st}\,.
\end{equation}
For $a<b<c$ we have
\begin{equation}
D_{n;st}^{abc}=
\begin{cases}\ang{a}{b}
\langle n t s |c\rangle
&\qquad b<s\leq c<t\\
\ang{a}{b}\ang{c}{n}x_{st}^2&\qquad b<s<t\leq c\\
\langle n t s |a\rangle
\ang{c}{b}&\qquad a<s\leq b,c<t\\
\ang{n}{a}\ang{b}{c}x_{st}^2&\qquad a<s<t\leq b\\
\ang{a}{b}\ang{c}{n}x_{st}^2
-\ang{a}{c}
\langle n t s |b\rangle
&\qquad a<s\leq b<t\leq c\\
\ang{a}{b}
\langle n s t |c\rangle
&\qquad s\leq a,b<t\leq c\\
\ang{c}{b}
\langle n s t |a\rangle
&\qquad s\leq a<t\leq b \,,
\end{cases}
\end{equation}
and $D^{abc}_{n;st}$ is totally antisymmetric in $a,b,c$.
Thus, the ${\mathcal N}=4$ super Yang-Mills NMHV six-fermion amplitude is
\begin{equation}
\left(A_n\right)^{\text{NMHV}}_{(q\bar{q})^3}=\frac{\delta^{(4)}(p)\sign(\tau)}{\ang{1}{2}\dots \ang{n}{1}}\sum_{1<s<t<n}\tilde{R}_{n;st}D_{n;st}^{\bar{b}_1\bar{b}_2}\begin{aligned}[t]\Bigl(&+\delta^{\scriptscriptstyle A_1B_1}\delta^{\scriptscriptstyle A_2B_2}\delta^{\scriptscriptstyle A_3B_3} D_{n;st}^{a_1\bar{b}_2}D_{n;st}^{\bar{b}_1a_2}D_{n;st}^{\bar{b}_1\bar{b}_2a_3}\\ &-\delta^{\scriptscriptstyle A_1B_2}\delta^{\scriptscriptstyle A_2B_1}\delta^{\scriptscriptstyle A_3B_3} D_{n;st}^{a_2\bar{b}_2}D_{n;st}^{\bar{b}_1a_1}D_{n;st}^{\bar{b}_1\bar{b}_2a_3}\\ &-\delta^{\scriptscriptstyle A_1B_3}\delta^{\scriptscriptstyle A_2B_2}\delta^{\scriptscriptstyle A_3B_1} D_{n;st}^{a_3\bar{b}_2}D_{n;st}^{\bar{b}_1a_2}D_{n;st}^{\bar{b}_1\bar{b}_2a_1}\\ &-\delta^{\scriptscriptstyle A_1B_1}\delta^{\scriptscriptstyle A_2B_3}\delta^{\scriptscriptstyle A_3B_2} D_{n;st}^{a_1\bar{b}_2}D_{n;st}^{\bar{b}_1a_3}D_{n;st}^{\bar{b}_1\bar{b}_2a_2}\\ &+\delta^{\scriptscriptstyle A_1B_2}\delta^{\scriptscriptstyle A_2B_3}\delta^{\scriptscriptstyle A_3B_1} D_{n;st}^{a_3\bar{b}_2}D_{n;st}^{\bar{b}_1a_1}D_{n;st}^{\bar{b}_1\bar{b}_2a_2}\\ &+\delta^{\scriptscriptstyle A_1B_3}\delta^{\scriptscriptstyle A_2B_1}\delta^{\scriptscriptstyle A_3B_2} D_{n;st}^{a_2\bar{b}_2}D_{n;st}^{\bar{b}_1a_3}D_{n;st}^{\bar{b}_1\bar{b}_2a_1}\Bigr) \, \end{aligned}
\end{equation}
which in the single-flavor case~\eqref{masterquark} reduces to
\begin{equation}
\left(A_n\right)^{\text{NMHV}}_{(q\bar{q})^3}
=\frac{\delta^{(4)}(p)\sign(\tau)}{\ang{1}{2}\dots \ang{n}{1}}
\sum_{1<s<t<n}\tilde{R}_{n;st}
\left(D_{n;st}^{\bar{b}_1\bar{b}_2}\right)^3D_{n;st}^{a_1a_2a_3}\,.
\end{equation}
These simplified expressions are implemented in {\tt GGT} by the functions
{\tt GGTnmhv6fermS} for the single-flavor case and 
{\tt GGTnmhv6ferm} for the general-flavor case.
See appendix \ref{GGT} for the documentation.

\subsection{\texorpdfstring{N${}^2$MHV}{NNMHV} amplitudes}
\subsubsection{Two fermions}
We continue the list of quark-gluon amplitudes by applying the master
formulas \eqref{masterquark} and \eqref{master_quark_SYM_original} 
in the N${}^2$MHV case with up to six fermions. The amplitude with
three negative-helicity gluons at positions $c_1$, $c_2$, $n$, 
a quark at position $a$ and an anti-quark at position $\bar{b}$, is
\begin{equation}
\left(A_n\right)_{(q\bar{q})}^{\text{N${}^2$MHV}}=
\frac{\delta^{(4)}(p)\sign(\tau)}{\ang{1}{2}\ldots\ang{n}{1}}
\sum_{2\leq a_1<b_1<n}
{\tilde R}_{n;a_1b_1}\begin{aligned}[t]\cdot\Biggl[
&\sum_{a_1+1\leq a_2<b_2\leq b_1}
{\tilde R}^{0;a_1 b_1}_{n;b_{1}a_{1};a_{2}b_{2}} 
\cdot D^{c_1c_2 a}_{\!\scriptscriptstyle1}
\left(D^{c_1c_2 \bar{b}}_{\!\scriptscriptstyle1}\right)^{3}\\
&+\sum_{b_{1}\leq a_{2}<b_{2}<n} {\tilde R}^{a_{1}b_{1};0}_{n;a_{2}b_{2}}
\cdot D^{c_1c_2 a}_{\!\scriptscriptstyle2}
\left(D^{c_1c_2 \bar{b}}_{\!\scriptscriptstyle2}\right)^{3}\, \Biggr]\,,
\end{aligned}
\end{equation}
with the $3\times3$ determinants $D^{abc}_{\!\scriptscriptstyle1}$
and $D^{abc}_{\!\scriptscriptstyle2}$ from eqs.~\eqref{eq:D1} and
\eqref{eq:D2}.
\subsubsection{Four fermions}
For the amplitude with two negative-helicity gluons at positions $c$, $n$,
as well as quarks and anti-quarks at positions
$\alpha_1^{A_1},\,\alpha_2^{A_2}$ and
$\bar{\beta}_1^{B_1},\,\bar{\beta}_2^{B_2}$, we obtain
\begin{multline}
\left(A_n\right)_{(q\bar{q})^2}^{\text{N${}^2$MHV}}=
\frac{\delta^{(4)}(p)\sign(\tau)}{\ang{1}{2}\ldots\ang{n}{1}}
\sum_{2\leq a_1<b_1<n}
{\tilde R}_{n;a_1b_1}\times\\
\times\begin{aligned}[t]\Biggl[&\sum_{a_1< a_2<b_2\leq b_1}
{\tilde R}^{0;a_1 b_1}_{n;b_{1}a_{1};a_{2}b_{2}} 
 \left(D^{c\bar{\beta}_1 \bar{\beta}_2}_{\!\scriptscriptstyle1}\right)^{2}
\left(\delta_{\scriptscriptstyle A_1}^{\scriptscriptstyle B_1}
\delta_{\scriptscriptstyle A_2}^{\scriptscriptstyle B_2}
D^{c\alpha_1 \bar{\beta}_2}_{\!\scriptscriptstyle1}
D^{c\bar{\beta}_1 \alpha_2}_{\!\scriptscriptstyle1}
-\delta_{\scriptscriptstyle A_1}^{\scriptscriptstyle B_2}
\delta_{\scriptscriptstyle A_2}^{\scriptscriptstyle B_1}
D^{c\alpha_2 \bar{\beta}_2}_{\!\scriptscriptstyle1}
D^{c\bar{\beta}_1 \alpha_1}_{\!\scriptscriptstyle1}\right)\\
&+\sum_{b_{1}\leq a_{2}<b_{2}<n} {\tilde R}^{a_{1}b_{1};0}_{n;a_{2}b_{2}}
\left(D^{c\bar{\beta}_1 \bar{\beta}_2}_{\!\scriptscriptstyle2}\right)^{2}
\left(\delta_{\scriptscriptstyle A_1}^{\scriptscriptstyle B_1}
\delta_{\scriptscriptstyle A_2}^{\scriptscriptstyle B_2}
D^{c\alpha_1 \bar{\beta}_2}_{\!\scriptscriptstyle2}
D^{c\bar{\beta}_1 \alpha_2}_{\!\scriptscriptstyle2}
-\delta_{\scriptscriptstyle A_1}^{\scriptscriptstyle B_2}
\delta_{\scriptscriptstyle A_2}^{\scriptscriptstyle B_1}
D^{c\alpha_2 \bar{\beta}_2}_{\!\scriptscriptstyle2}
D^{c\bar{\beta}_1 \alpha_1}_{\!\scriptscriptstyle2}\right)
\, \Biggr]\end{aligned} 
\end{multline}
in the $\mathcal{N}=4$ super Yang-Mills case, and
\begin{equation}
\left(A_n\right)_{(q\bar{q})^2}^{\text{N${}^2$MHV}}=
\frac{\delta^{(4)}(p)\sign(\tau)}{\ang{1}{2}\ldots\ang{n}{1}}
\sum_{2\leq a_1<b_1<n}
{\tilde R}_{n;a_1b_1}\begin{aligned}[t]\cdot\Biggl[&\sum_{a_1< a_2<b_2\leq b_1}
{\tilde R}^{0;a_1 b_1}_{n;b_{1}a_{1};a_{2}b_{2}} 
\cdot D^{c\alpha_1\alpha_2}_{\!\scriptscriptstyle1}
\left(D^{c\bar{\beta}_1 \bar{\beta}_2}_{\!\scriptscriptstyle1}\right)^{3}\\
&+\sum_{b_{1}\leq a_{2}<b_{2}<n} {\tilde R}^{a_{1}b_{1};0}_{n;a_{2}b_{2}}
\cdot D^{c_1c_2 a}_{\!\scriptscriptstyle2}
\left(D^{c_1c_2 \bar{b}}_{\!\scriptscriptstyle2}\right)^{3}\, \Biggr]
\end{aligned}
\end{equation}
for single-flavor QCD.

\subsubsection{Six fermions}
For the $\mathcal{N}=4$ super Yang-Mills amplitude with one
negative-helicity gluon at position $n$, quarks and anti-quarks at
positions $\alpha_1^{A_1}\!,\,\alpha_2^{A_2}\!,\,\alpha_3^{A_3}$ and
$\bar{\beta}_1^{B_1}\!,\,\bar{\beta}_2^{B_2}\!,\,\bar{\beta}_3^{B_3}$,
our master formula yields
\begin{multline}
\left(A_n\right)_{(q\bar{q})^3}^{\text{N${}^2$MHV}}=
\frac{\delta^{(4)}(p)\sign(\tau)}{\ang{1}{2}\ldots\ang{n}{1}}
\sum_{2\leq a_1<b_1<n}
{\tilde R}_{n;a_1b_1}\times\\
\times\!\begin{aligned}[t]\Biggl[&\sum_{a_1< a_2<b_2\leq b_1}
\!{\tilde R}^{0;a_1 b_1}_{n;b_{1}a_{1};a_{2}b_{2}} \,
 D^{\bar{\beta}_1 \bar{\beta}_2\bar{\beta}_3}_{\!\scriptscriptstyle1}
\left(\delta_{\scriptscriptstyle A_{1}}^{\scriptscriptstyle B_1}
\delta_{\scriptscriptstyle A_{2}}^{\scriptscriptstyle B_2}
\delta_{\scriptscriptstyle A_{3}}^{\scriptscriptstyle B_3}
D^{\alpha_{1} \bar{\beta}_2\bar{\beta}_2}_{\!\scriptscriptstyle1}
D^{ \bar{\beta}_1\alpha_{2}\bar{\beta}_2}_{\!\scriptscriptstyle1}
D^{ \bar{\beta}_2\bar{\beta}_2\alpha_{3}}_{\!\scriptscriptstyle1}
\pm 
\text{permutations of $\;\begin{Bmatrix}A_i\\\alpha_i\end{Bmatrix}$}\right)\\
&+\sum_{b_{1}\leq a_{2}<b_{2}<n} \!{\tilde R}^{a_{1}b_{1};0}_{n;a_{2}b_{2}}\,
 D^{\bar{\beta}_1 \bar{\beta}_2\bar{\beta}_3}_{\!\scriptscriptstyle2}
\left(\delta_{\scriptscriptstyle A_{1}}^{\scriptscriptstyle B_1}
\delta_{\scriptscriptstyle A_{2}}^{\scriptscriptstyle B_2}
\delta_{\scriptscriptstyle A_{3}}^{\scriptscriptstyle B_3}
D^{\alpha_{1} \bar{\beta}_2\bar{\beta}_2}_{\!\scriptscriptstyle2}
D^{ \bar{\beta}_1\alpha_{2}\bar{\beta}_2}_{\!\scriptscriptstyle2}
D^{ \bar{\beta}_2\bar{\beta}_2\alpha_{3}}_{\!\scriptscriptstyle2}
\pm 
\text{permutations of $\;\begin{Bmatrix}A_i\\\alpha_i\end{Bmatrix}$}\right)
\, \Biggr]\,,\end{aligned}
\end{multline}
which in the single-flavor case simplifies to
\begin{equation}
\left(A_n\right)_{(q\bar{q})^3}^{\text{N${}^2$MHV}}=
\frac{\delta^{(4)}(p)\sign(\tau)}{\ang{1}{2}\ldots\ang{n}{1}}\sum_{2\leq a_1<b_1<n}
{\tilde R}_{n;a_1b_1}\begin{aligned}[t]\cdot\Biggl[&\sum_{a_1< a_2<b_2\leq b_1}
{\tilde R}^{0;a_1 b_1}_{n;b_{1}a_{1};a_{2}b_{2}} \,
D^{\alpha_1\alpha_2\alpha_3}_{\!\scriptscriptstyle1}\left(D^{\bar{\beta}_1\bar{\beta}_2\bar{\beta}_3}_{\!\scriptscriptstyle1}\right)^{3}\\
&+\sum_{b_{1}\leq a_{2}<b_{2}<n} {\tilde R}^{a_{1}b_{1};0}_{n;a_{2}b_{2}}\,
D^{\alpha_1\alpha_2\alpha_3}_{\!\scriptscriptstyle2}\left(D^{\bar{\beta}_1\bar{\beta}_2\bar{\beta}_3}_{\!\scriptscriptstyle2}\right)^{3}
\, \Biggr] \,. \end{aligned}
\end{equation}
We recall that these formulas hold for arbitrary color-orderings of the 
$n$ partons.

We have implemented all of the above simplified expressions for
NNMHV amplitudes with up to six fermions in the functions
{\tt GGTnnmhv2ferm}, {\tt GGTnnmhv4ferm}, {\tt GGTnnmhv6ferm} in the
{\tt GGT} package.

\section{The Mathematica package {\tt GGT}}
\label{GGT}

Here we describe the Mathematica package {\tt GGT} (gluon-gluino trees)
provided with the {\tt arXiv.org} submission of the present paper and
also accessible via {\tt http://qft.physik.hu-berlin.de}.\\

The idea is to provide the formulas derived in the present paper in
computer-readable form, such that the interested reader can use them
without having to type them in. We have also included a simple numerical 
evaluation routine for given phase-space points in the {\tt GGT} package,
as well as an interface to the
spinor-helicity package {\tt S@M} \cite{Maitre:2007jq}. 
The issue of computer speed optimization will be commented upon below.\\

Let us now describe the different functions in {\tt GGT} and then give a
specific example.
The following functions are provided in {\tt GGT}
\begin{itemize}
\item {\tt GGTgluon[n,H]} \\
gives the $n$-gluon amplitude (\ref{masterglue}), with the positions of
the negative-helicity gluons given by the list {\tt H}.
\item {\tt GGTfermionS[n, gluonlist, fermlist, afermlist]}\\
gives the $n$-parton amplitude (\ref{masterquark})
of an arbitrary number of gluons and single-flavor fermion/antifermions. 
The positions of the
negative-helicity gluons, helicity $+\frac12$ fermions, and helicity
$-\frac12$ anti-fermions are given by the lists {\tt gluonlist},
{\tt fermlist}, and {\tt afermlist}, respectively.
\item {\tt GGTfermion[n, gluonlist, fermlist, afermlist]}\\
is the generalization of {\tt GGTfermionS} to multiple fermion flavors, eq. (\ref{master_quark_SYM_original}).
The positions of the
negative-helicity gluons
are given by the list {\tt gluonlist}.
The positions $q_{i}, \bar{q}_{i}$ and flavors $A_{i}, B_{i}$ of the helicity $+\frac12$ fermions
and helicity $-\frac12$ anti-fermions are given by the lists 
{\tt fermlist}$=\{ \{ q_{i}, A_{i} \}, \ldots  \}$, and {\tt afermlist}$=\{ \{ \bar{q}_{i}, A_{i} \}, \ldots  \}$, respectively.
\item {\tt GGTsuperamp[$n,k$]}\\
is the N${}^k$MHV superamplitude of $n$ superfields, with the 
MHV superamplitude factored out, in terms of the $R$ invariants.
\end{itemize}
Let us give an example.  We can load the {\tt GGT} package using
\begin{flushleft}
\begin{tt}
<< GGT.m
\end{tt}
\end{flushleft}
Suppose we want to evaluate a gluon amplitude.
Typing
\begin{flushleft}
\begin{tt}
GGTgluon[6,\{3,5,6\}]
\end{tt}
\end{flushleft}
prints the $6$-gluon NMHV amplitude with
helicity configuration ${+}{+}{-}{+}{-}{-}$,
\begin{align}\nonumber
& \frac{1   
   }{\langle 1|2\rangle 
   \langle 2|3\rangle  \langle 3|4\rangle  \langle 4|5\rangle  \langle 5|6\rangle  \langle 6|1\rangle } \\
& \bigg( \frac{\langle 2|1\rangle  \langle 4|3\rangle  \left(s_{2,4} \langle 6|3\rangle  \langle 6|5\rangle +\left\langle 6|x_{6,4}|x_{4,2}|3\right\rangle 
   \langle 6|5\rangle \right){}^4}{s_{2,4} \left\langle 6|x_{6,2}|x_{2,4}|3\right\rangle  \left\langle 6|x_{6,2}|x_{2,4}|4\right\rangle  \left\langle
   6|x_{6,4}|x_{4,2}|1\right\rangle  \left\langle 6|x_{6,4}|x_{4,2}|2\right\rangle } \nonumber \\
   &+\frac{\langle 2|1\rangle  \langle 5|4\rangle  \left(s_{2,5} \langle
   6|3\rangle  \langle 6|5\rangle 
       \left\langle 6|x_{6,5}|x_{5,2}|3\right\rangle  \langle 6|5\rangle \right){}^4}{s_{2,5} \left\langle
   6|x_{6,2}|x_{2,5}|4\right\rangle  \left\langle 6|x_{6,2}|x_{2,5}|5\right\rangle  \left\langle 6|x_{6,5}|x_{5,2}|1\right\rangle  \left\langle
   6|x_{6,5}|x_{5,2}|2\right\rangle } \nonumber \\
   &+\frac{\langle 3|2\rangle  \langle 5|4\rangle  \left(s_{3,5} \langle 6|3\rangle  \langle 6|5\rangle +\left\langle
   6|x_{6,5}|x_{5,3}|3\right\rangle  \langle 6|5\rangle \right){}^4}{s_{3,5} \left\langle 6|x_{6,3}|x_{3,5}|4\right\rangle  \left\langle
   6|x_{6,3}|x_{3,5}|5\right\rangle  \left\langle 6|x_{6,5}|x_{5,3}|2\right\rangle  \left\langle 6|x_{6,5}|x_{5,3}|3\right\rangle } \bigg) \nonumber
\end{align}
{\tt GGT} formatted the output for better readability. The underlying
formula, which can be accessed explicitly, {\it e.g.}~by using 
{\tt Inputform[...]}, depends on the following quantities:  The spinor
products $\langle i j \rangle$ are denoted by {\tt GGTspaa[i,j]}.
Differences between dual coordinates
$x_{i,j} = p_{i} + p_{i+1} + \ldots + p_{j-1}$
are denoted by {\tt GGTx[i,j] }. Finally, the
abbreviation $x_{ij}^2 = s_{i,j-1}$ is used and denoted by 
{\tt GGTs[i,j-1]}.

In order to obtain numerical values,
we can use the spinor-helicity package {\tt S@M} \cite{Maitre:2007jq}. 
The function {\tt GGTtoSpinors} converts the expression
into one that can be evaluated by the latter package.
In our example, the commands
\begin{flushleft}
\begin{tt}
<< Spinors.m\\
GenMomenta[1,2,3,4,5,6] 
\end{tt}
\end{flushleft}
load the {\tt S@M} package and use one of its functions to generate
arbitrary momenta for a six-particle scattering process.
Finally, numerical values of the amplitude at that phase-space point can
be obtained by the command
\begin{flushleft}
\begin{tt}
GGTtoSpinors[GGTgluon[6,\{3,5,6\}]] //N
\end{tt}
\end{flushleft}
A faster implementation for the numerical
evaluation of the {\tt GGT} formulas is provided by the 
function {\tt GGTgenvar[P]} which generates the spinors and region
momenta for a numerical evaluation of an amplitude at a desired
phase-space point $P=\{p_{1},p_{2},\ldots,p_{n}\}$.
For example, for the kinematic point given in eq.~(4.6) of
ref.~\cite{Ellis:2006ss} (which to save space we give here to only
three significant digits), one would use
\begin{flushleft}
\begin{tt}
GGTgenvar[$
\{\{
-3.0, 2.12,  1.06,  1.84 \}$, $\{
-3.0,-2.12, -1.06, -1.84 \}$, $\{
 2.0, 2.0,   0.0 ,  0.0 \}$, $\{
 0.857, -0.316, 0.797, 0.0 \}$, $\{
 1.0, -0.184, 0.465, 0.866 \}$, $\{
 2.14, -1.5, -1.26, -0.866 \}\}
 $]
\end{tt}
\end{flushleft}
One can then evaluate an amplitude numerically by the command
\begin{flushleft}
\begin{tt}
GGTnumeric[GGTfermionS[$6, \{1, 6\}, \{2, 4\}, \{3, 5\}$]] \\[0.1cm]
- 0.496838 + 0.0714737 i
\end{tt}
\end{flushleft}
%
%
This approach is considerably faster than the 
{\tt GGTtoSpinors[\ldots]//N} function discussed above.

Let us comment about the evaluation time needed using our approach.
It is clear that for any serious applications or for comparisons with other
methods, one should implement our analytical formulas using a 
low-level programming language, such as C, C++ or FORTRAN.
For example, an implementation of the NMHV formulas in C++ 
results in a speedup of orders of magnitude over a similar 
implementation in Mathematica.
Moreover, it is important to efficiently cache (store the numerical
values of) quantities that are used
repeatedly. In this spirit, the Mathematica demonstration package {\tt GGT}
provides a computer-readable version of the formulas needed
for such an approach, so that the user does not have to type them in
manually.

Our analytical formulas are very similar, and in some cases identical,
to the ones obtained in a very recent paper~\cite{Bourjaily:2010wh}.
The latter also correspond to solutions of the BCFW recursion
relations, based
on refs.~\cite{ArkaniHamed:2009si,Bourjaily:2010kw}, but may differ
in form since they can correspond to different factorization channels.
Another difference is that they are written using momentum-twistor
variables~\cite{Hodges:2009hk,Mason:2009qx}. Ref.~\cite{Bourjaily:2010wh}
contains a numerical Mathematica implementation of these formulas.
When the formulas of our paper and that of ref.~\cite{Bourjaily:2010wh}
are both implemented with appropriate caching in C++, for the NMHV
tree amplitudes for $Vq\bar{q}ggggg$ and $Vq\bar{q}Q\bar{Q}ggg$,
their evaluation time is similar~\cite{privateZviHarald}.

We remark that in approaches based on BCFW recursion relations, the asymptotic
number of terms in N${}^{k}$MHV${}_{n}$ amplitudes as $n$ becomes large
is quadratic in $n$ for NMHV, quartic for NNMHV and worse for higher $k$.
This is the reason we especially simplified the NMHV and NNMHV formulas
presented in our paper, since we expect that they will be the most useful
for practical applications, especially for small $n$. For $k > 2$ and 
large $n$ there are at least two efficient numerical strategies making
use of these formulae.  First, one could use our formulae as initial
conditions for a numerical implementation  of the BCFW recursion
relations, as described in section 3.  Alternatively,
one could use the Berends-Giele approach for $k > 2$, implemented using an
efficient caching, in combination with our formulas for
$k \le 2$~\cite{privateUwerBiedermann}.

We also included further functions that evaluate directly the simplified 
amplitudes of Appendix~\ref{explicit} and \ref{explicit-fermions}.
They can be accessed via the following functions.
\begin{itemize}
\item {\tt GGTnmhvgluon[$n,a,b$]}\\ 
is the simplified $n$-parton NMHV gluon amplitude
with negative-helicity gluons at positions $a,b$ and $n$.
\item {\tt GGTnnmhvgluon[$n,a,b,c$]}\\ 
is the simplified $n$-parton NNMHV gluon 
amplitude with negative-helicity gluons at positions $a,b,c$ and $n$.
\item {\tt GGTnmhv2ferm[$n,c,a,\bar{b}$]}\\
is the simplified $n$-parton NMHV two-fermion amplitude with 
negative-helicity gluons at positions $c, n$ and a fermion/anti-fermion 
at positions $a$ and $\bar{b}$.
\item {\tt GGTnnmhv2ferm[$n,c_{1},c_{2},a,\bar{b}$]}\\
is the simplified $n$-parton
NNMHV two-fermion amplitude with negative-helicity gluons at 
positions $c_{1},c_{2}$ and $n$ and a 
fermion/anti-fermion at position $a$ and $\bar{b}$. 
\item {\tt GGTnmhv4ferm[$n, \{ \{a_{1},A_{1}\} ,\{a_{2},A_{2}\}\} , \{ \{\bar{b}_{1},B_{1}\} ,\{\bar{b}_{2},B_{2}\}\}$]}\\
is the simplified $n$-parton NMHV four-fermion amplitude
with a negative-helicity gluon at position $n$, two gluinos of flavors $A_{i}$ at positions $a_{i}$ and
two anti-gluinos of flavors $B_{i}$ at positions $\bar{b}_{i}$.
\item {\tt GGTnmhv4fermS[$n, \{ a_{1} ,a_{2}\} , \{ \bar{b}_{1},\bar{b}_{2}\}$]}\\
is the simplified $n$-parton NMHV four-fermion amplitude
with a negative-helicity gluon at position $n$ and equally flavored gluinos/anti-gluinos at positions
$a_{i}$ and $\bar{b}_{i}$, respectively.
\item {\tt GGTnnmhv4ferm[$n,c, \{ \{a_{1},A_{1}\} ,\{a_{2},A_{2}\}\} , \{ \{\bar{b}_{1},B_{1}\} ,\{\bar{b}_{2},B_{2}\}\}$]}\\
 is the simplified $n$-parton NNMHV four-fermion amplitude
with two negative-helicity gluons at position $c,n$, two gluinos of 
flavors $A_{i}$ at positions $a_{i}$ and
two anti-gluinos of flavors $B_{i}$ at positions $\bar{b}_{i}$.
\item {\tt GGTnnmhv4fermS[$n,c, \{ a_{1} ,a_{2}\} , \{ \bar{b}_{1},\bar{b}_{2}\}$]}\\
is the simplified $n$-parton NNMHV four-fermion amplitude
with negative-helicity gluons at positions $c,n$ and equally flavored gluinos/anti-gluinos at positions
$a_{i}$ and $\bar{b}_{i}$, respectively.
\item {\tt GGTnmhv6ferm[$n, B_3 ,  \{ \{a_{1},A_{1}\} ,\{a_{2},A_{2}\},\{a_{3},A_{3}\}\} , \{ \{\bar{b}_{1},B_{1}\} ,\{\bar{b}_{2},B_{2}\}\}$]}\\
 is the simplified $n$-parton NMHV six-fermion amplitude with three 
gluinos of flavors $A_{i}$ at positions $a_{i}$ and 
three anti-gluinos of flavors $B_{i}$ at positions 
$\bar{b}_{i}$. Note that $\bar{b}_{3}=n$.
\item {\tt GGTnmhv6fermS[$n,   \{ a_{1} , a_{2}, a_{3}\} , \{ \bar{b}_{1},\bar{b}_{2}\}$]}\\
 is the simplified $n$-parton NMHV six-fermion amplitude with equally flavored gluinos/anti-gluinos at positions $a_{i}$ and 
$\bar{b}_{i}$, respectively. Note that $\bar{b}_{3}=n$.
\item {\tt GGTnnmhv6ferm[$n,  \{ \{a_{1},A_{1}\} ,\{a_{2},A_{2}\},\{a_{3},A_{3}\}\} , \{ \{\bar{b}_{1},B_{1}\} ,\{\bar{b}_{2},B_{2}\},\{\bar{b}_{3},B_{3}\}\}$]}\\
is the simplified $n$-parton NNMHV six-fermion amplitude with a negative helicity gluon
at position $n$ and three 
gluinos of flavors $A_{i}$ at positions $a_{i}$ and 
three anti-gluinos of flavors $B_{i}$ at positions 
$\bar{b}_{i}$.
\item {\tt GGTnnmhv6fermS[$n, \{ a_{1} , a_{2}, a_{3}\} , \{ \bar{b}_{1},\bar{b}_{2},
\bar{b}_{3}\}$]}\\
is the simplified $n$-parton NNMHV six-fermion amplitude with a negative helicity gluon
at position $n$ and equally flavored gluinos/anti-gluinos at positions $a_{i}$ and 
$\bar{b}_{i}$, respectively.
\end{itemize}
The full list of functions available in {\tt GGT} can be
obtained by typing
\begin{flushleft}
\begin{tt}
\$GGTfunctions 
\end{tt}
\end{flushleft}
along with the documentation of each implemented function 
that can be accessed via the command
\begin{flushleft}
\begin{tt}
?GGTgluon
\end{tt}
\end{flushleft}
for example.


\bibliographystyle{nb}
\bibliography{botany}

\end{document}